\documentclass[twocolappendix]{emulateapj}
\usepackage{graphicx}
\usepackage{amssymb}
\usepackage{amsmath}
\usepackage{natbib}
\usepackage{color}
\newcommand\be{\begin{equation}}
\newcommand\ee{\end{equation}}

\newcommand\ba{\begin{eqnarray}}
\newcommand\ea{\end{eqnarray}}

\begin{document}

\title{Planet-disk interaction in disks with cooling: basic theory}

\author{Ryan Miranda\altaffilmark{1,3} and Roman R. Rafikov\altaffilmark{1,2}}

\altaffiltext{1}{Institute for Advanced Study, Einstein Drive, Princeton, NJ 08540}
\altaffiltext{2}{Centre for Mathematical Sciences, Department of Applied Mathematics and Theoretical Physics, University of Cambridge, Wilberforce Road, Cambridge CB3 0WA, UK}
\altaffiltext{3}{miranda@ias.edu}

\begin{abstract}
Gravitational coupling between young planets and their parent disks is often explored using numerical simulations, which typically treat the disk thermodynamics in a highly simplified manner. In particular, many studies adopt the locally isothermal approximation, in which the disk temperature is a fixed function of the stellocentric distance. We explore the dynamics of planet-driven density waves in disks with more general thermodynamics, in which the temperature is relaxed towards an equilibrium profile on a finite cooling timescale $t_{\rm c}$. We use both linear perturbation theory and direct numerical simulations to examine the global structure of density waves launched by planets in such disks. A key diagnostic used in this study is the behavior of the wave angular momentum flux (AMF), which directly determines the evolution of the underlying disk. The AMF of free waves is constant for slowly cooling (adiabatic) disks, but scales with the disk temperature for rapidly cooling (and locally isothermal) disks. However, cooling must be extremely fast, with $\beta = \Omega t_{\rm c} \lesssim 10^{-3}$ for the locally isothermal approximation to provide a good description of density wave dynamics in the linear regime (relaxing to $\beta \lesssim 10^{-2}$ when nonlinear effects are important). For intermediate cooling timescales, density waves are subject to a strong linear damping. This modifies the appearance of planet-driven spiral arms and the characteristics of axisymmetric structures produced by massive planets: in disks with $\beta \approx 0.1$ -- $1$, a near-thermal mass planet opens only a single wide gap around its orbit, in contrast to the several narrow gaps produced when cooling is either faster or slower.
\end{abstract}


\section{Introduction}
\label{sect:intro}

The gravitational interaction of a gaseous disk with a massive orbital companion plays an important role in many astrophysical systems, including circumstellar (i.e. protoplanetary) disks, cataclysmic variables, and disk galaxies. The tidal gravitational potential of the companion excites density waves at Lindblad resonances---locations in the disk at which the natural frequency of the disk is commensurate with the forcing frequency of the companion \citep{GT80}. These waves then travel across the disk carrying angular momentum and energy with them over large distances. Their  dissipation, either due to linear damping (e.g., viscous damping, \citealt{Takeuchi1996}), or nonlinear dissipation \citep{GR01,R02}, leads to the deposition of the wave angular momentum into the background disk fluid and completes the process of the global angular momentum transport in the disk \citep{Lunine1982,GN89,RP12}. This transfer of angular momentum from the density waves after their dissipation can be a significant driver of disk evolution \citep{GR01,R16,AR18}, often resulting in the formation of axisymmetric features such as gaps and rings \citep{R02b,DongGaps2017}. The detailed outcome of such planet-disk coupling depends critically on the angular momentum flux (AMF) carried by the waves. As shown by \citet{GT79}, in adiabatic disks the AMF of free waves (i.e., not subject to external torques) is conserved in the linear regime and in the absence of dissipation. However, for other thermodynamic assumptions the AMF behavior may change.

In numerical simulations of protoplanetary disks, the disk thermodynamics are often treated in a highly simplified manner by using the so-called locally isothermal approximation. In this approximation, the sound speed $c_\mathrm{s}$, or equivalently the disk temperature $T$, is assumed to be a prescribed function of the radial coordinate $r$ only, dispensing with the need to solve an energy equation for the disk gas. In \citet{Miranda-ALMA}, we showed that in the locally isothermal disks, the AMF of free waves is {\it not conserved}, in contrast to the adiabatic disks studied in \citet{GT79}. Instead, AMF is proportional to $c_\mathrm{s}^2$. Since typically the disk temperature decreases with radius, this means that waves traveling inward {\it accumulate AMF} as they propagate. This occurs as a result of {\it extracting} angular momentum from the background disk flow (so that the total angular momentum of the disk-wave system is conserved), an effect previously pointed out by \citet{Lin2011} and \citet{Lin2015}. This has important consequences for wave-driven disk evolution, since the impact of a (dissipating) wave on the disk gets enhanced by this AMF amplification process as the wave propagates to smaller and smaller radii in locally isothermal disks \citep{Miranda-ALMA}. 

Adiabatic and locally isothermal disks represent the extreme limits of a more general thermodynamics, in which the disk temperature is relaxed towards an equilibrium profile on a finite timescale. Physically, the locally isothermal approximation corresponds to the scenario in which (1) the imposed temperature profile is maintained externally, e.g., by irradiation from the central star, and (2) deviations from the imposed temperature profile, associated with either compression/expansion of the gas or radial displacement of fluid elements, are quickly neutralized by the radiation or absorption of thermal energy. Here ``quickly'' means that the timescale for erasing temperature perturbations, which we refer to loosely as the cooling timescale $t_\mathrm{c}$, is small compared to all other relevant timescales. On the contrary, the adiabatic limit is expected apply when $t_\mathrm{c}$ is very long. A common---but not rigorously motivated---assumption is that the relevant timescale separating these two limits and to which $t_\mathrm{c}$ should be compared is the orbital timescale.

In this paper, we carry out a linear perturbation analysis for disks with thermodynamics affected by thermal relaxation (cooling), in order to understand the behavior of planet-excited density waves and implications for wave-driven disk evolution. An important result of this analysis is the derivation of a ``master equation'' describing the global behavior of non-axisymmetric perturbations driven by an external gravitational potential in a two-dimensional (2D) disk. Such an equation was first presented by \cite{GT79}, for the case of adiabatic perturbations in disks with uniform entropy, and was later generalized to disks with general entropy profiles \citep{Baruteau2008,Tsang2014}. We present an even more general version of the equation for adiabatic perturbations in disks with thermal relaxation, which reduces to the locally isothermal and adiabatic regimes in the appropriate limits (short and long cooling timescales, respectively). By solving for the perturbations excited by an embedded planet, we determine the full, global structure of planet-driven density waves (e.g., \citealt{OL02,Miranda-Spirals}). These calculations are corroborated using fully nonlinear numerical simulations of low-mass planets in disks with cooling.

Another key result of this paper is the analysis of the behavior of the angular momentum flux (AMF) for free waves in disks with cooling/thermal relaxation in the linear regime. We not only confirm that the AMF behavior reduces to the adiabatic limit (i.e., conserved) for sufficiently long $t_\mathrm{c}$, and to the locally isothermal limit (i.e., proportional to $c_\mathrm{s}^2$) for sufficiently short $t_\mathrm{c}$, but also {\it quantify} the conditions on the cooling timescale required for these regimes to be realized. In particular, we show that the condition required for the locally isothermal approximation to be valid is much more stringent than expected from simply requiring that the cooling timescale $t_\mathrm{c}$ is smaller than the orbital timescale. Instead, $t_\mathrm{c}$ must be an very small fraction ($\lesssim 10^{-3}$) of the orbital timescale. We also show that cooling leads to a strong linear damping of density waves for a range of cooling timescales. This has significant consequences for disk evolution driven by density waves.

The plan for this paper is as follows. In Section~\ref{sect:theory}, we present the linear analysis for non-axisymmetric perturbations driven by an external potential in disks with different thermodynamic assumptions, deriving the master equation describing the global structure of perturbations in each case. We numerically validate the linear analysis using numerical simulations in Section~\ref{sect:validation}. In Section~\ref{sect:amf}, we analyze the behavior of the wave AMF under the different thermodynamic assumptions, including a discussion of the disk torques. In Section~\ref{sect:massive}, we explore the role of cooling on disk evolution driven by a massive planet. We discuss our results in Section~\ref{sect:discussion} and summarize our main conclusions in Section \ref{eq:sum}.

\section{Linear Perturbation Theory with Different Thermodynamic Assumptions}
\label{sect:theory}

In this section, we derive different versions of the master equation for non-axisymmetric linear perturbations driven by an external potential in two-dimensional disks. We consider three different thermodynamic assumptions: (i) adiabatic perturbations in disks with radially-varying entropy profiles (``adiabatic disks''), (ii) isothermal perturbations in disks with fixed radial temperature profiles (``locally isothermal disks''), and (iii) perturbations in disks in which the internal energy (or temperature) is relaxed towards a prescribed profile on a finite cooling timescale (``disks with cooling''). We progress in a pedagogical fashion to highlight the differences arising due to varying thermodynamic assumptions. In each case, we discuss the reduction to the master equation of \citet{GT79} for adiabatic perturbations in uniform entropy disks.

\subsection{Basic Assumptions and General Approach}

We consider an inviscid two-dimensional gas disk that is subject to an external potential. The disk is described in polar coordinates $(r,\phi)$ by the surface density $\Sigma$, height-integrated pressure $P = c_\mathrm{s,iso}^2 \Sigma$ (where $c_\mathrm{s,iso}=(k_\mathrm{B}T/\mu)^{1/2}$ is the isothermal sound speed), radial velocity $u_r$, and azimuthal velocity $u_\phi$. The unperturbed disk is axisymmetric and described by the density $\Sigma_0(r)$, pressure $P_0(r)$, radial velocity $u_{r,0}(r) = 0$ and azimuthal velocity $u_{\phi,0}(r) = r\Omega(r)$, where $\Omega(r)$ is the rotation frequency. We consider perturbations to the background state, $\Sigma = \Sigma_0 + \delta\Sigma, P = P_0 + \delta P, u_r = u_{r,0} + \delta u_r$, and $u_\phi = u_{\phi,0} + \delta u_\phi$. For convenience, we will typically drop the subscripts from the unperturbed variables. The perturbed quantities are assumed to have the form of Fourier harmonics, i.e.,
\be
\delta x(r,\phi,t) = \delta x(r) \exp[\mathrm{i}m(\phi-\omega_\mathrm{p} t)],
\ee
for any perturbed variable $\delta x$. Here $\omega_\mathrm{p}$ is the pattern frequency of the perturbation. The perturbed variables satisfy the following dynamical (mass and momentum conservation) equations:
\begin{gather}
\label{eq:pert1}
-\mathrm{i}\tilde{\omega}\delta\Sigma + \frac{1}{r}\frac{\partial}{\partial r}(r\Sigma\delta u_r) + \frac{\mathrm{i}m\Sigma}{r} \delta u_\phi = 0, \\
\label{eq:pert2}
-\mathrm{i}\tilde{\omega}\delta u_r - 2\Omega\delta u_\phi = -\frac{1}{\Sigma}\frac{\partial}{\partial r} \delta P + \frac{1}{\Sigma^2} \frac{\mathrm{d}P}{\mathrm{d}r} \delta\Sigma - \frac{\partial}{\partial r} \Phi_m, \\
\label{eq:pert3}
-\mathrm{i}\tilde{\omega} \delta u_\phi + \frac{\kappa^2}{2\Omega} \delta u_r = -\frac{\mathrm{i}m}{r}\left(\frac{\delta P}{\Sigma} + \Phi_m\right).
\end{gather}
Here $\tilde{\omega} = m(\omega_\mathrm{p} - \Omega)$ is the Doppler-shifted frequency of the perturbation, $\Phi_m$ is the Fourier component of the external potential which has azimuthal number $m$ and rotates at the rate $\omega_\mathrm{p}$, and $\kappa^2 = (2\Omega/r)(r^2\Omega)^\prime$ is the squared radial epicyclic frequency (the prime denotes the radial derivative).

In order to provide a full description of behavior of the perturbations, equations~(\ref{eq:pert1})--(\ref{eq:pert3}) must be supplemented with an equation of state (EoS), which relates $P$ to $\Sigma$, as well as to other thermodynamic quantities. Specifying an EoS provides a fourth perturbation equation relating $\delta P$ to $\delta\Sigma$ (and potentially other fluid variables). This equation, along with equations~(\ref{eq:pert1})--(\ref{eq:pert3}), then form a closed system. Through algebraic substitution, these equations can be combined into a single equation, or master equation, for one variable only. 

We choose as the preferred variable the ``enthalpy'' perturbation $\delta h = \delta P/\Sigma$. Note that $\delta h$ is strictly equal to the true thermodynamic enthalpy perturbation only in isentropic disks. Nonetheless, the variable $\delta h$ defined in this way serves as a convenient variable for which a master equation can be found.

Several useful intermediate results in the derivation of the final master equation for $\delta h$ for each of the different thermodynamic assumptions are given in Appendix~\ref{sect:pert-eqns} and \ref{sect:velocity-pert}. In Appendix~\ref{sect:pert-eqns}, equations (\ref{eq:pert1})--(\ref{eq:pert3}) are expressed with $\delta\Sigma$ and $\delta P$ eliminated in favor of $\delta h$ only, and in Appendix~\ref{sect:velocity-pert}, the velocity perturbations $\delta u_r$ and $\delta u_\phi$ are given in terms of the enthalpy perturbation $\delta h$ and its radial derivative. The velocity perturbations expressed in this way are also useful for the analysis of AMF conservation presented in Section~\ref{sect:amf}.

\subsection{Adiabatic Disks}
\label{sect:theory-adi}

The master equation for planet-driven waves in an adiabatic, non-barotropic disk (i.e., with a radially-varying entropy $S$) has been derived previously \citep{Baruteau2008,Tsang2014}. For completeness, we briefly restate its derivation here.

We assume an ideal equation of state
\be
P = (\gamma-1)e\Sigma,
\ee
where $\gamma$ is the adiabatic index and $e$ is the specific internal energy, which is related to the adiabatic sound speed of the disk according to $c_\mathrm{s,adi}^2 = \gamma(\gamma - 1) e = \gamma c_\mathrm{s,iso}^2$.
For adiabatic perturbations, the fluid entropy $S \propto \ln(P/\Sigma^\gamma)$ is conserved in the Lagrangian sense, i.e., $\mathrm{d}S/\mathrm{d}t = 0$. This results in the energy equation for the total (background $+$ perturbation) $P$ and $\Sigma$,
\be
\label{eq:eng}
\frac{\mathrm{d}e}{\mathrm{d}t} + P\frac{\mathrm{d}}{\mathrm{d}t}\left(\frac{1}{\Sigma}\right) = 0.
\ee
The energy equation for the perturbed fluid variables is therefore
\be
\label{eq:eng_adi}
-\mathrm{i}\tilde{\omega}(\delta P - c_\mathrm{s,adi}^2\delta\Sigma) = -\frac{\Sigma c_\mathrm{s,adi}^2}{L_S}\delta u_r.
\ee
Here we have defined
\be
\label{eq:LS}
\frac{1}{L_S} = \frac{1}{\gamma}\frac{\mathrm{d} S}{\mathrm{d}r},
\ee
the inverse length scale of entropy variation.

Equations~(\ref{eq:pert1})--(\ref{eq:pert3}) and (\ref{eq:eng_adi}) are combined into a single second-order equation for $\delta h$, which we represent in the form
\be
\label{eq:master}
\frac{\mathrm{d}^2}{\mathrm{d}r^2}\delta h + C_1\frac{\mathrm{d}}{\mathrm{d}r}\delta h + C_0\delta h = \Psi_m,
\ee
where
\begin{align}
\label{eq:c1_adi}
C_1 & = \frac{\mathrm{d}}{\mathrm{d}r} \ln\left(\frac{r\Sigma}{D_S}\right), \\
\begin{split}
\label{eq:c0_adi}
C_0 & = -\frac{2m\Omega}{r\tilde{\omega}}\left[\frac{2}{L_S} + \frac{\mathrm{d}}{\mathrm{d}r}\ln\left(\frac{\Sigma\Omega}{D_S}\right)\right] - \frac{1}{L_S^2} \\
& - \frac{1}{L_S}\frac{\mathrm{d}}{\mathrm{d}r}\ln\left(\frac{r\Sigma}{L_S D_S}\right) - \frac{m^2}{r^2}\left(1 - \frac{N_r^2}{\tilde{\omega}^2}\right) - \frac{D_S}{c_\mathrm{s,adi}^2},
\end{split}
\end{align}
and the forcing due to the planetary potential is 
\be
\label{eq:psi_adi}
\begin{gathered}
\Psi_m = -\frac{\mathrm{d}^2\Phi_m}{\mathrm{d}r^2} - \left[\frac{1}{L_S} + \frac{\mathrm{d}}{\mathrm{d}r} \ln\left(\frac{r\Sigma}{D_S}\right)\right] \frac{\mathrm{d}\Phi_m}{\mathrm{d}r} \\
 + \left\{\frac{2m\Omega}{r\tilde{\omega}}\left[\frac{\mathrm{d}}{\mathrm{d}r}\ln\left(\frac{\Sigma\Omega}{D_S}\right) + \frac{1}{L_S}\right] \right. \\
\left. + \frac{m^2}{r^2}\left(1 - \frac{N_r^2}{\tilde{\omega}^2}\right) \right\} \Phi_m.
\end{gathered}
\ee
Here
\be
\label{eq:DS}
D_S = \kappa^2 - \tilde{\omega}^2 + N_r^2,
\ee
and
\be
\label{eq:Nr2}
N_r^2 = -\frac{1}{\Sigma^2}\frac{\mathrm{d}P}{\mathrm{d}r}\left(\frac{1}{c_\mathrm{s,adi}^2}\frac{\mathrm{d}P}{\mathrm{d}r}-\frac{\mathrm{d}\Sigma}{\mathrm{d}r}\right)
\ee
is the square of the Brunt--V\"{a}is\"{a}l\"{a} frequency. 

For barotropic disks ($S =$ constant), $L_S \rightarrow \infty$ and $N_r \rightarrow 0$, resulting in the reduction of equations (\ref{eq:c1_adi})--(\ref{eq:psi_adi}) to the \citet{GT79} equation.

\subsection{Locally Isothermal Disks}
\label{sect:theory-iso}

In the locally isothermal treatment of the disk thermodynamics, a fixed temperature profile $T(r)$ is assumed. This assumption corresponds to the situation in which $T(r)$ is set by, e.g., irradiation by the central star, and in which radiative cooling eliminates any temperature variations very quickly. As a result, $c_\mathrm{s,iso}^2$ is a fixed function of $r$, eliminating the need for an explicit energy equation. The locally isothermal approximation has been widely used in numerical simulations, but to the best of our knowledge, the full master equation for planet-driven waves has not been formulated before (\citealt{Lee2016} has previously derived only the homogeneous part of the equation).

As a result of making the locally isothermal assumption, the EoS (in terms of the total $P$ and $\Sigma$) is expressed as
\be
P = c_\mathrm{s,iso}^2(r) \Sigma,
\ee
and corresponding equation for the perturbed variables is simply
\be
\label{eq:eng_iso}
\delta P = c_\mathrm{s,iso}^2(r) \delta \Sigma.
\ee
Combining equations (\ref{eq:pert1})--(\ref{eq:pert3}) and (\ref{eq:eng_iso}), we find a master equation for $\delta h$ (see equation~(\ref{eq:master})):
\begin{align}
\label{eq:c1_iso}
C_1 & = \frac{\mathrm{d}}{\mathrm{d}r} \ln\left(\frac{r\Sigma}{D}\right) - \frac{1}{L_T}, \\
\begin{split}
\label{eq:c0_iso}
C_0 & = -\frac{2m\Omega}{r\tilde{\omega}} \left[\frac{1}{L_T} + \frac{\mathrm{d}}{\mathrm{d}r}\ln\left(\frac{\Sigma\Omega}{D}\right)\right] \\
& - \frac{1}{L_T}\frac{\mathrm{d}}{\mathrm{d}r}\ln \left(\frac{r\Sigma}{L_T D}\right) - \frac{m^2}{r^2} - \frac{D}{c_\mathrm{s,iso}^2},
\end{split}
\end{align}
and the forcing is
\be
\label{eq:psi_iso}
\begin{aligned}
\Psi_m = & -\frac{\mathrm{d}^2\Phi_m}{\mathrm{d}r^2} - \left[\frac{\mathrm{d}}{\mathrm{d}r}\ln\left(\frac{r\Sigma}{D}\right)\right] \frac{\mathrm{d}\Phi_m}{\mathrm{d}r} \\
& + \left\{\frac{2m\Omega}{r\tilde{\omega}}\left[\frac{\mathrm{d}}{\mathrm{d}r}\ln\left(\frac{\Sigma\Omega}{D}\right)\right] + \frac{m^2}{r^2}\right\} \Phi_m.
\end{aligned}
\ee
Here we have defined
\be
\label{eq:D}
D = \kappa^2 - \tilde{\omega}^2,
\ee
and
\be
\label{eq:LT}
\frac{1}{L_T} = \frac{\mathrm{d}\ln c_\mathrm{s}^2}{\mathrm{d}r},
\ee
the length scale of the variation of $T \propto c_\mathrm{s}^2$ (here $c_\mathrm{s}$ refers to either adiabatic or isothermal sound speed). 

Note that if we take the limit $\gamma \rightarrow 1$ in the adiabatic case (\S \ref{sect:theory-adi}), then $D_S \rightarrow D$, $L_S \rightarrow L_T$ and $N_r\rightarrow 0$. However, the expressions (\ref{eq:c1_adi})--(\ref{eq:psi_adi}) do not reduce to their locally isothermal analogues (\ref{eq:c1_iso})--(\ref{eq:psi_iso}) in this limit. This observation highlights the {\it singular nature} of the locally isothermal approximation and clearly shows why this approximation leads to different results when compared to the adiabatic case with $\gamma\rightarrow 1$, something that has been pointed out in \citet{Miranda-ALMA}.

Only in the {\it globally} isothermal limit ($ T=$ const), when $L_T \rightarrow \infty$, equations (\ref{eq:c1_adi})--(\ref{eq:psi_adi}) become identical to the equations~(\ref{eq:c1_iso})--(\ref{eq:psi_iso}) and reduce to the master equation of \citet{GT79}.

\subsection{Disks with Cooling}
\label{sect:theory-cool}

The adiabatic and locally isothermal disks considered in the previous subsections represent limiting cases of a more general disk thermodynamics, in which the disk temperature is relaxed towards an equilibrium profile on a finite timescale. We now derive the generalized master equation for disks with cooling.

Analogous to adiabatic disks, we adopt an ideal equation of state. We add a cooling term on the right hand side of equation~(\ref{eq:eng}), which relaxes $e$ towards a prescribed equilibrium profile $e_0(r) = c_\mathrm{s,adi}^2(r)/[\gamma(\gamma-1)]$ on a cooling\footnote{The energy source term described by equation~(\ref{eq:cooling_law}) represents both heating and cooling of the gas toward a fixed temperature. Therefore it represents thermal relaxation rather than strictly cooling. However, we will nonetheless loosely refer to it as ``cooling''.} timescale $t_\mathrm{c}$:
\be
\label{eq:cooling_law}
\left(\frac{\partial e}{\partial t}\right)_\mathrm{cool} = -\frac{e-e_0}{t_\mathrm{c}}.
\ee
Note that we allow $t_\mathrm{c}$ to be an arbitrary function of $r$. Equations (\ref{eq:eng}) and (\ref{eq:cooling_law}) lead to the energy equation for the perturbed fluid variables,
\be
\label{eq:eng_cool}
\left(\frac{1}{t_\mathrm{c}}-\mathrm{i}\tilde{\omega}\right)\delta P - \left(\frac{1}{\gamma t_\mathrm{c}}-\mathrm{i}\tilde{\omega}\right) c_\mathrm{s,adi}^2 \delta\Sigma = -\frac{\Sigma c_\mathrm{s,adi}^2}{L_S}\delta u_r.
\ee
Note that equation~(\ref{eq:eng_cool}) reduces to corresponding equation (\ref{eq:eng_iso}) for locally isothermal disks in the limit $t_\mathrm{c} \rightarrow 0$ (noting that $c_\mathrm{s,adi}^2/\gamma = c_\mathrm{s,iso}^2$), and to the corresponding equation (\ref{eq:eng_adi}) for adiabatic disks in the limit $t_\mathrm{c} \rightarrow \infty$.

Combining equations (\ref{eq:pert1})--(\ref{eq:pert3}) and (\ref{eq:eng_cool}), we find that the master equation for the general cooling case is (see equation~(\ref{eq:master})):
\begin{align}
\label{eq:c1_cool}
C_1 & = \frac{\mathrm{d}}{\mathrm{d}r} \ln\left(\frac{r\Sigma}{D_\mathrm{c}}\right) - \frac{1}{(1-\mathrm{i}\gamma\tilde{\beta})L_T}, \\
\begin{split}
\label{eq:c0_cool}
C_0 & = -\frac{2m\Omega}{r\tilde{\omega}} \left[\left(\frac{L_T^{-1}-2\mathrm{i}\gamma\tilde{\beta}L_S^{-1}}{1-\mathrm{i}\gamma\tilde{\beta}}\right) + \frac{\mathrm{d}}{\mathrm{d}r}\ln\left(\frac{\Sigma\Omega}{D_\mathrm{c}}\right)\right] \\
& + \frac{\gamma\tilde{\beta}L_S^{-1}(\gamma\tilde{\beta}L_S^{-1}+\mathrm{i}L_T^{-1})}{(1-\mathrm{i}\gamma\tilde{\beta})^2} \\
& - \left(\frac{L_T^{-1}-\mathrm{i}\gamma\tilde{\beta}L_S^{-1}}{1-\mathrm{i}\gamma\tilde{\beta}}\right)\frac{\mathrm{d}}{\mathrm{d}r}\ln\left[\frac{r\Sigma}{D_\mathrm{c}}\left(\frac{L_T^{-1}-\mathrm{i}\gamma\tilde{\beta}L_S^{-1}}{1-\mathrm{i}\gamma\tilde{\beta}}\right)\right] \\
& - \left(\frac{\kappa^2-D_\mathrm{c}}{\tilde{\omega}^2}\right)\frac{m^2}{r^2} - \left(\frac{1-\mathrm{i}\tilde{\beta}}{1-\mathrm{i}\gamma\tilde{\beta}}\right)\frac{\gamma D_\mathrm{c}}{c_\mathrm{s,adi}^2},
\end{split}
\end{align}
with forcing due to the planetary potential 
\be
\label{eq:psi_cool}
\begin{gathered}
\Psi_m = -\frac{\mathrm{d}^2\Phi_m}{\mathrm{d}r^2} + \left[\frac{\mathrm{i}\gamma\tilde{\beta}}{(1-\mathrm{i}\gamma\tilde{\beta})L_S} - \frac{\mathrm{d}}{\mathrm{d}r} \ln\left(\frac{r\Sigma}{D_\mathrm{c}}\right)\right] \frac{\mathrm{d}\Phi_m}{\mathrm{d}r} \\
+ \Bigg\{\frac{2m\Omega}{r\tilde{\omega}}\left[\frac{\mathrm{d}}{\mathrm{d}r}\ln\left(\frac{\Sigma\Omega}{D_\mathrm{c}}\right) - \frac{\mathrm{i}\gamma\tilde{\beta}}{(1-\mathrm{i}\gamma\tilde{\beta})L_S}\right] \\
+ \left(\frac{\kappa^2-D_\mathrm{c}}{\tilde{\omega}^2}\right)\frac{m^2}{r^2} \Bigg\} \Phi_m.
\end{gathered}
\ee
Here $\tilde{\beta} = \tilde{\omega}t_\mathrm{c}$ and
\be
\label{eq:Dc}
D_\mathrm{c} = D - \frac{\mathrm{i}\gamma\tilde{\omega}t_\mathrm{c}}{1-\mathrm{i}\gamma\tilde{\omega}t_\mathrm{c}} N_r^2.
\ee
It is easily verified that equations (\ref{eq:c1_cool})--(\ref{eq:psi_cool}) reduce to the master equation for adiabatic disks (equations (\ref{eq:c1_adi})--(\ref{eq:psi_adi})) for $t_\mathrm{c} \rightarrow \infty$, and to the master equation for locally isothermal disks (equations (\ref{eq:c1_iso})--(\ref{eq:psi_iso})) for $t_\mathrm{c} \rightarrow 0$, as expected. Note that the coefficients $C_1$ and $C_0$ (as well as the coefficients of $\Phi_m$ in the source term $\Psi_m$) are real in these limits, but are complex in general for a finite $t_\mathrm{c}$. 

\section{Numerical Validation}
\label{sect:validation}

\begin{figure*}
\begin{center}
\includegraphics[width=0.99\textwidth,clip]{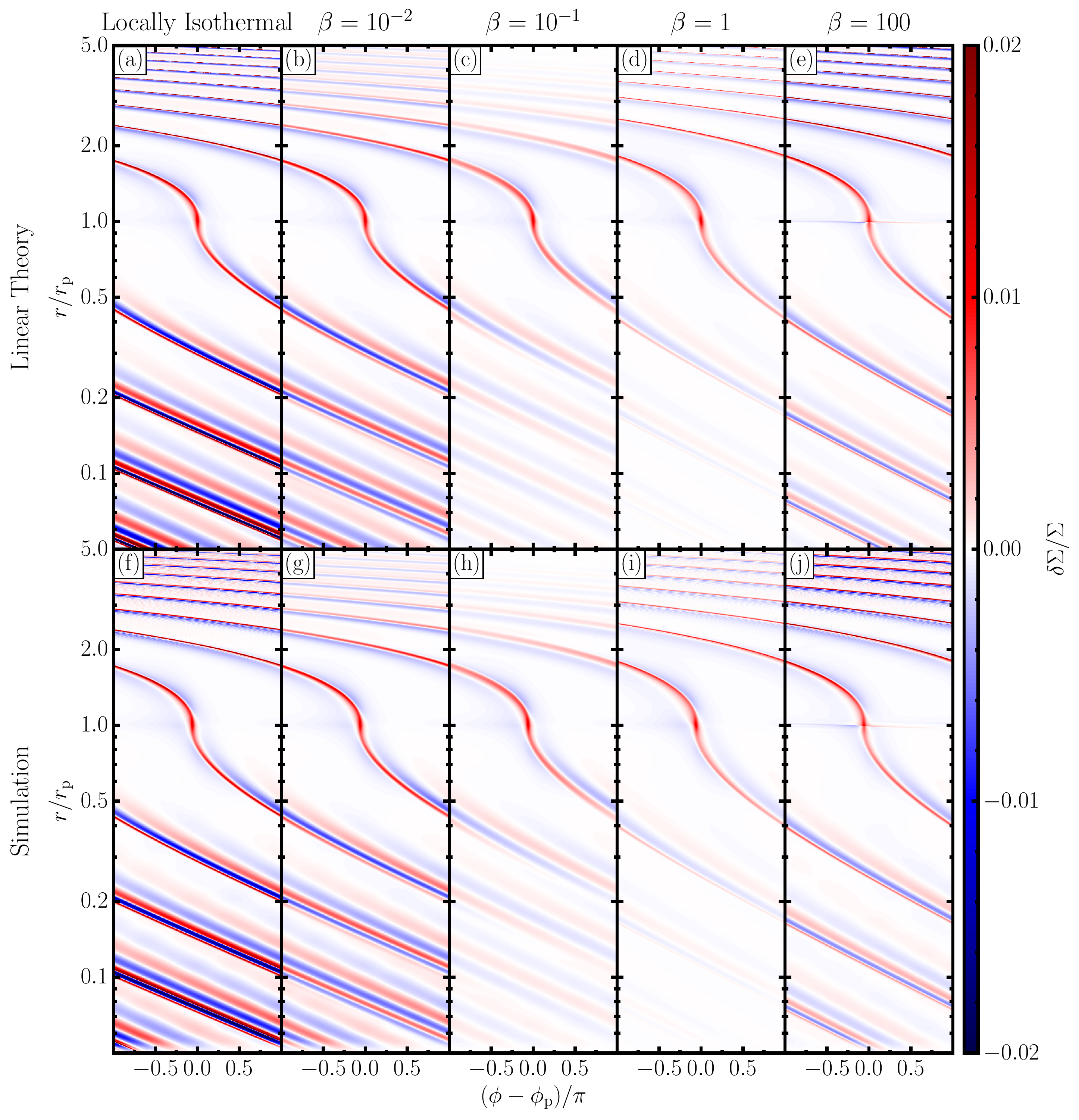}
\caption{Fractional surface density perturbation $\delta\Sigma/\Sigma$, shown in polar coordinates for disks with different thermodynamics: locally isothermal ((a), (f)), and with different constant values of the dimensionless cooling timescale $\beta = \Omega t_\mathrm{c}$ (all other panels). The top row ((a)--(e)) shows the results of numerical solutions of the linear perturbation equations ((\ref{eq:master}), (\ref{eq:c1_iso})--(\ref{eq:psi_iso}) in panel (a) and (\ref{eq:master}), (\ref{eq:c1_cool})--(\ref{eq:psi_cool}) in other panels), and the bottom row ((f)--(j)) shows the results of nonlinear simulations with a $0.01 M_\mathrm{th}$ planet (for which the perturbation is well in the linear regime) at $20$ orbits. The disk has an aspect ratio $h_\mathrm{p} = 0.1$, temperature power law index $q = 1$, surface density power law $p = 1$, and (for the non-locally isothermal calculations) adiabatic index $\gamma = 7/5$.}
\label{fig:spirals}
\end{center}
\end{figure*}

The master equation (\ref{eq:master}), (\ref{eq:c1_cool})--(\ref{eq:psi_cool}) can be used to analytically determine the details of the density wave excitation and their subsequent propagation in the linear regime, as has been done in a number of past studies  \citep{GT79,Zhang2006,Tsang2014}. However, in this work, we take a different approach, making use of numerical solutions of the master equation (following e.g., \citealt{KP93,RP12}) to characterize the waves excited by low-mass planets in protoplanetary disks. 

We consider the interaction of a disk around a star of mass $M_*$ with a planet of mass $M_\mathrm{p} \ll M_*$. The planet is assumed to have a circular orbit with radius $r_\mathrm{p}$ and orbital period $t_\mathrm{p} = 2\pi/\Omega_\mathrm{p}$, where $\Omega_\mathrm{p} = (GM_*/r_\mathrm{p}^3)^{1/2}$. We present numerical solutions of the different versions of the master equation for planet-driven waves derived before, and describe the basic global structure of the perturbations. We also present the results of direct numerical simulations of low-mass planets in disks, for which the disk response is well-approximated by the linear regime, in order to validate our linear analysis.

\subsection{Disk Model}

The analysis presented in Section~\ref{sect:theory} is valid for disks with arbitrary profiles for the disk surface density $\Sigma(r)$ and temperature $T(r)$. For the purposes of the following numerical calculations it is necessary to adopt a concrete disk model, and so we choose a simple power law disk. The unperturbed disk has an {\it isothermal} sound speed (see Section~\ref{sect:cs}) given by
\be
c_\mathrm{s,iso}(r) = h_\mathrm{p} r_\mathrm{p}\Omega_\mathrm{p} \left(\frac{r}{r_\mathrm{p}}\right)^{-q/2},
\label{eq:cs}
\ee
where $h_\mathrm{p}$ is the disk aspect ratio, $h(r) = H/r = h_\mathrm{p}(r/r_\mathrm{p})^{(1-q)/2}$, at $r_p$. Here $H = c_\mathrm{s,iso}/\Omega$ is the pressure scale height. The parameter $q$ is the power law index of the disk temperature $T\propto c_\mathrm{s,iso}^2$. The surface density profile is
\be
\Sigma(r) = \Sigma_\mathrm{p}\left(\frac{r}{r_\mathrm{p}}\right)^{-p},
\ee
where the value of $\Sigma_\mathrm{p}$ is arbitrary and $p$ is a constant. Accounting for the radial pressure support, the orbital frequency and radial epicyclic frequency resulting from centrifugal balance are therefore
\begin{align}
\Omega & = \Omega_\mathrm{K} [1 - h^2(r)(q+p)]^{1/2}, \\
\kappa & = \Omega_\mathrm{K} [1 - h^2(r)(q+p)(2-q)]^{1/2},
\end{align}
where $\Omega_\mathrm{K} = (GM_*/r^3)^{1/2}$ is the Keplerian frequency. The (inverse) lengthscales of the variation of entropy and temperature (equations (\ref{eq:LS}) and (\ref{eq:LT})) are
\be
\frac{1}{L_S} = \frac{(\gamma-1)p-q}{\gamma r}, \quad \frac{1}{L_T} = -\frac{q}{r}.
\ee

We choose the disk aspect ratio at the location of the planet $h_\mathrm{p} = 0.1$, and temperature power law index $q = 1$. As a result, the aspect ratio $h(r)$ is independent of $r$ (this a theoretical convenience, and is not necessarily intended to describe realistic disks). We choose a surface density power law index $p = 1$, and (except in our locally isothermal calculations) adiabatic index $\gamma = 7/5$.

\subsection{Linear Calculations}
\label{sect:lin-calc}

We compute the linear response of the disk to an orbiting planet by solving the master equation for either disks with cooling (equation~(\ref{eq:master}) with $C_1, C_0,$ and $\Psi_m$ given by equations~(\ref{eq:c1_cool})--(\ref{eq:psi_cool})) or for locally isothermal disks (in which case $C_1, C_0,$ and $\Psi_m$ are given by equations~(\ref{eq:c1_iso})--(\ref{eq:psi_iso})), for modes with different azimuthal numbers. The solution method, which closely follows that of \citet{KP93}, is described in detail in the appendix of \citet{Miranda-Spirals}.\footnote{In setting the outgoing wave boundary conditions using the asymptotic (WKB) behavior of solutions, the variation of the wave amplitude is modified as appropriate based on which version of the master equation is being solved. The amplitude variation follows from the behavior of the wave AMF, see Section~\ref{sect:amf}.} We solve for the structure of modes with $m \le m_\mathrm{max}$, where $m_\mathrm{max} = 80$ is sufficient to achieve convergence of the perturbation structure. We then construct the full two-dimensional structure of the perturbed fluid variables by synthesizing them in real space.

For planets on circular orbits, the Fourier harmonics of the gravitational potential are
\be
\Phi_m = -\frac{GM_\mathrm{p}}{r_\mathrm{p}} b_{1/2}^{(m)}(r/r_\mathrm{p}),
\ee
where
\be
b_{1/2}^{(m)}(\alpha) = \frac{1}{\pi} \int_0^{2\pi} \frac{\cos(m\psi)\mathrm{d}\psi}{\left[1 - 2\alpha\cos(\psi) + \alpha^2 + \epsilon^2\right]^{1/2}}
\ee
are softened Laplace coefficients. We choose for the softening parameter $\epsilon = 0.6 h_\mathrm{p}$, corresponding the softening length of $0.6 H_\mathrm{p}$ (representing the effect of the finite vertical extent of the disk). Note that for a circular orbit, the wave pattern frequency $\omega_\mathrm{p}$ is equal to the orbital frequency of the planet $\Omega_\mathrm{p}$ for all $m$. We ignore the indirect potential term, $\delta_{m,1}GM_\mathrm{p}r/r_\mathrm{p}^2$, associated with the orbital motion of the central star, as it has a negligible impact on the overall perturbation structure \citep{Miranda-Spirals}.

\subsection{Hydrodynamical Simulations}
\label{sect:sims}

We also run 2D inviscid hydrodynamical simulations of planet-disk interaction using {\sc fargo3d} \citep{FARGO3d}. The numerical grid extends from $r_\mathrm{in} = 0.05 r_\mathrm{p}$ to $r_\mathrm{out} = 5.0 r_\mathrm{p}$ with logarithmic spacing in the radial direction and uniform spacing in the azimuthal direction. The number of grid cells is $N_r \times N_\phi = 3004 \times 4096$. Wave damping \citep{deValBorro2006} is applied near the inner and outer boundaries ($r < 0.06 r_\mathrm{p}$ and $r > 4.5 r_\mathrm{p}$) to prevent wave reflection. The mass of the planet is gradually increased from zero to $M_\mathrm{p}$ over $10$ orbits, and its potential is softened with a lengthscale $0.6 H_\mathrm{p}$, as in the linear calculations.

We choose either a locally isothermal EoS, or an ideal EoS with $\gamma = 7/5$ and cooling (equation~(\ref{eq:cooling_law})). The same initial temperature profile is used in both cases, in order to provide the most direct comparison. Cooling is implemented using a simple implicit (backward Euler) step performed after the main hydro step. This implementation is stable and performs as expected (based on agreement with linear theory) for cooling timescales at least as small as $10^{-4} \Omega^{-1}$.

We choose a planet mass $M_\mathrm{p} = 10^{-5} M_* = 0.01 M_\mathrm{th}$, where
\be
M_\mathrm{th} = h_\mathrm{p}^3 M_*
\ee
is the thermal mass. Since $M_\mathrm{p} \ll M_\mathrm{th}$, the response of disk to the planet is well-approximated by the linear regime---at least close to the planet. The waves may undergo significant nonlinear evolution and develop into shocks at large distances from the planet. However, as described in \citet{Miranda-Spirals}, the waves excited by a $0.01 M_\mathrm{th}$ planet never develop into shocks in the inner disk, as a result of the partitioning of the wave AMF into multiple spiral arms (cf.~\citealt{R02}). Shocks do develop at large radii in the outer disk, where there is only one spiral arm carrying all of the angular momentum flux.

\subsection{Structure of Perturbations}
\label{sect:pert_struct}

Fig.~\ref{fig:spirals} shows the disk surface density perturbation, $\delta\Sigma$ (in terms of the background surface density $\Sigma$) for a planet in a locally isothermal disk and in disks with different dimensionless cooling timescales $\beta = \Omega t_\mathrm{c}$. The chosen values of $\beta$, $10^{-2}$ -- $10^2$, along with the locally isothermal case, representing the limit $\beta \rightarrow 0$, represent the full range of behaviors for the different thermodynamics we consider. The results of both numerical solutions of the linear master equation (Fig.~\ref{fig:spirals}(a)--(e)) and numerical simulations (Fig.~\ref{fig:spirals}(f)--(j)) are shown. The simulation results show excellent agreement with the linear results, down to fine details, confirming the validity of our linear analysis, as well as of our numerical implementation of cooling in the simulations.

Generically, the perturbations take the form of a single spiral arm (i.e., an azimuthally narrow maximum of $\delta\Sigma$) in the outer disk, and multiple spirals (as many as four) gradually emerging in the inner disk. The spirals are the result of the complicated interference of perturbations with different azimuthal mode numbers, which can be understood based on the dispersion relation for linear density waves \citep{BZ18a,Miranda-Spirals}. This basic mechanism is evidently not strongly affected by the intricacies of the disk thermodynamics. However, the detailed structure of the spirals is affected by a particular choice of the non-zero cooling timescale in several key ways.

The first effect is that the radial range of the waves depends strongly on the cooling timescale. For fast cooling ($\beta \lesssim 10^{-2}$; Fig.~\ref{fig:spirals}(a)--(b)) or slow cooling (e.g., $\beta = 100$; Fig.~\ref{fig:spirals}(e)), the waves propagate far into both the inner and outer disk. That is, the wave amplitude at radii far from the planet is comparable to (or in some cases larger than) the wave amplitude near the planet. However, for the cases with $\beta = 10^{-1}$ and $\beta = 1$ (Fig.~\ref{fig:spirals}(c),(d)), the wave amplitude decreases with distance from the planet. The waves are most strongly damped in the inner disk for the case $\beta = 1$ (Fig.~\ref{fig:spirals}(d)), for which the amplitude decreases (relative to the amplitude close to the planet) by nearly an order of magnitude by the time the waves reach $0.5 r_\mathrm{p}$. In the outer disk, waves are most strongly damped when the cooling time is somewhat shorter, e.g., the case with $\beta= 10^{-1}$ (Fig.~\ref{fig:spirals}(c)). In this case, the wave amplitude has decreased by an order of magnitude when the waves reach $2.5 r_\mathrm{p}$.

A second effect is that the the overall tightness or pitch angle of the spirals (i.e., slope of density contours in Fig.~\ref{fig:spirals}) varies with the value of $\beta$. The spirals are most tightly wound for very small values of $\beta$, and least tightly wound for large values of $\beta$. This is related to the dependence of the effective sound speed for density waves on the cooling timescale, see Section~\ref{sect:discussion-spirals} for a more in-depth discussion. 

Finally, the details of the structure of the multiple spirals in the inner disk---their relative amplitudes, separations, and azimuthal widths---are modified by cooling. Although we forgo a detailed analysis of the multiple spiral structure as carried out in \citet{Miranda-Spirals}, trends regarding the multiple spiral structure are evident in Fig~\ref{fig:spirals}. For $\beta = 10^{-1}$ -- $1$, the primary spiral arm (i.e., the one that touches the planet) is wider than it is for either smaller or larger values of $\beta$, and the primary and secondary spirals are more widely separated.  The case with $\beta = 10^{-1}$ represents an extreme modification of the spiral structure; far in the inner disk ($r \lesssim 0.1 r_\mathrm{p}$), the surface density is almost a pure $m = 2$ sinusoid. This pattern is very distinct from the narrow, azimuthally concentrated spiral arms found when $\beta$ is very large or very small; it will be discussed further in Section \ref{sect:discussion-spirals}. 

\section{Angular Momentum Flux}
\label{sect:amf}

We can gain insight into the behavior of density waves by considering the behavior of the wave angular momentum flux (AMF),
\be
F_J(r) = r^2 \Sigma(r) \oint \delta u_r (r,\phi) \delta u_\phi(r,\phi) \mathrm{d}\phi, 
\ee
where $\delta u_r$ and $\delta u_\phi = u_\phi - r\Omega$ are the velocity perturbations. For a Fourier mode with azimuthal number $m$, the mode AMF is given in terms of the complex perturbations $\delta u_r(r), \delta u_\phi(r) \propto \exp(\mathrm{i}m\phi)$ by
\be
\label{eq:fj_gen}
F_J^m = \pi r^2\Sigma~\mathrm{Re}(\delta u_r \delta u_\phi^*).
\ee
The total AMF $F_J$ is found by summing over all of the modes,
\be
F_J = \sum_{m=1}^\infty F_J^m.
\ee

We wish to describe the behavior of the AMF for free waves, i.e., not subject to an external potential. Evolution of the AMF of free waves is tied to wave-driven evolution of the disk (see Sections \ref{sect:massive} and \ref{sect:discussion-mdot}). Planet-excited waves can be considered free at locations far from the planet, where the torque density,
\be
\label{eq:dtdr}
\frac{\mathrm{d}T}{\mathrm{d}r} = -r\oint \Sigma \frac{\partial \Phi_\mathrm{p}}{\partial \phi} \mathrm{d}\phi,
\ee
which describes wave excitation, is negligible (here $\Phi_\mathrm{p}$ is the gravitational potential of the planet). Practically speaking, this is satisfied beyond about $2$ -- $3$ scale heights $H_\mathrm{p}$ from the planet \citep{Dong2011a,RP12}. 

Our goal is to derive a conservation law for the Fourier AMF $F_J^m$ of free waves for each of the thermodynamic assumptions described in Section~\ref{sect:theory}. The general strategy is to express $F_J^m$ (equation~(\ref{eq:fj_gen})) in terms of $\delta h$ (see Appendix~\ref{sect:velocity-pert}), so that the homogeneous version of the master equation (\ref{eq:master}) for $\delta h$ and either (\ref{eq:c1_adi})--(\ref{eq:c0_adi}), (\ref{eq:c1_iso})--(\ref{eq:c0_iso}), or (\ref{eq:c1_cool})--(\ref{eq:c0_cool}) (with $\Psi_m = 0$) can be used to ascertain the behavior of $F_J$.

\subsection{Adiabatic Disks}
\label{sect:amf-adi}

For adiabatic disks, plugging $\delta u_r$ and $\delta u_\phi$ given by equations (\ref{eq:ur_adi})--(\ref{eq:uphi_adi}) into the expression (\ref{eq:fj_gen}) for $F_J$ we find that the angular momentum flux for free waves is
\be
\label{eq:fj_adi}
F_J^m = \frac{\pi m r\Sigma}{D_S} \mathrm{Im}(\delta h \delta h^{*\prime}),
\ee
where the prime denotes the radial derivative. A global conservation law for $F_J$ can be found by taking the complex conjugate of the general homogeneous version of the master equation ((\ref{eq:master}) with $\Psi_m = 0$), multiplying by $\delta h$, and taking the imaginary part of the resulting equation, leading to
\be
\label{eq:master_cons}
\mathrm{Im}(\delta h \delta h^{*\prime\prime}) + C_1\mathrm{Im}(\delta h \delta h^{*\prime}) = 0.
\ee
On the other hand, differentiating (\ref{eq:fj_adi}) and making use of the expression (\ref{eq:c1_adi}) for $C_1$ for adiabatic disks, one can easily see that equation~(\ref{eq:master_cons}) is equivalent to
\be
\label{eq:fj_law_adi}
\frac{\mathrm{d}F_J^m}{\mathrm{d}r} = 0.
\ee
Therefore, $F_J^m$ is constant or conserved (i.e., independent of $r$) in adiabatic disks \citep{GT79}. Since the AMF of each wave mode is conserved, the {\it total} wave AMF must also be conserved:
\be
\label{eq:fj_law_adi_full}
\frac{\mathrm{d}F_J}{\mathrm{d}r} = 0.
\ee

\subsection{Locally Isothermal Disks}
\label{sect:amf-iso}

For locally isothermal disks, using expresions (\ref{eq:ur_iso})--(\ref{eq:uphi_iso}) for $\delta u_r$ and $\delta u_\phi$ in equation (\ref{eq:fj_gen}), we find that $F_J$ is given by
\be
\label{eq:fj_iso}
F_J^m = \frac{\pi m r\Sigma}{D} \mathrm{Im}(\delta h \delta h^{*\prime}),
\ee
i.e., the same as for adiabatic disks (equation~(\ref{eq:fj_adi})), but with $D_S \rightarrow D$. We follow the same procedure as for the case of adiabatic disks to find a global conservation law for $F_J$. In this case, $C_1$ is given by equation~(\ref{eq:c1_iso}). As a result, we see that equation~(\ref{eq:master_cons}) is equivalent to
\be
\label{eq:fj_law_iso}
\frac{\mathrm{d}}{\mathrm{d}r}\left(\frac{F_J^m}{c_\mathrm{s}^2}\right) = 0.
\ee
Therefore, $F_J^m$ is not constant, but instead proportional to $c_\mathrm{s}^2$ in locally isothermal disks (stated without a proof in \citealt{Lee2016}). Since equation~(\ref{eq:fj_law_iso}) applies to all wave modes, the total wave AMF obeys the same conservation law:
\be
\label{eq:fj_law_iso_full}
\frac{\mathrm{d}}{\mathrm{d}r}\left(\frac{F_J}{c_\mathrm{s}^2}\right) = 0,
\ee
generally disagreeing with equation (\ref{eq:fj_law_adi}) even when $\gamma\rightarrow 1$ in the latter. Only in the limit of a {\it globally} isothermal disk ($L_T \rightarrow \infty$) equation (\ref{eq:fj_law_iso_full}) reduces to equation (\ref{eq:fj_law_adi_full}).

\subsection{Disks with Cooling}
\label{sect:amf-cool}

For disks with cooling, using equations (\ref{eq:ur_cool})--(\ref{eq:uphi_cool}), we find that the Fourier AMF is given in terms of $\delta h$ by
\be
\label{eq:fj_cool}
\begin{aligned}
F_J^m & = \frac{\pi m r\Sigma}{D^2+\gamma^2\tilde{\beta}^2D_S^2} \left\{\vphantom{\frac{0}{0}}(D+\gamma^2\tilde{\beta}^2 D_S)\mathrm{Im}(\delta h\delta h^{*\prime}) \right. \\ 
& \left.- \gamma\tilde{\beta}N_r^2\mathrm{Re}(\delta h\delta h^{*\prime}) \right. \\ 
& \left. + \gamma\tilde{\beta}\left[\frac{2m\Omega}{r\tilde{\omega}}N_r^2 - \frac{D}{L_S} + \frac{D_S}{L_T}\right]|\delta h|^2\right\}.
\end{aligned}
\ee
In the limit $t_\mathrm{c} \rightarrow 0$, we recover equation~(\ref{eq:fj_iso}) and for $t_\mathrm{c} \rightarrow \infty$, we recover equation~(\ref{eq:fj_adi}).

The complexity of equations (\ref{eq:c1_cool})--(\ref{eq:psi_cool}) and (\ref{eq:fj_cool}) precludes us from finding a global conservation law for the Fourier AMF, as in the adiabatic and locally isothermal cases. Instead, we use a local WKB analysis to determine the approximate behavior of $F_J^m$ (e.g., \citealt{Takeuchi1996}), the details of which are given in Appendix~\ref{sect:wkb-cool}. The result is
\be
\label{eq:fj_law_cool}
\begin{aligned}
F_J^m(r) & = F_J^m(r_0) \\
& \times \exp\left\{\int_{r_0}^r \left[\frac{\Omega^2 L_T^{-1}}{\Omega^2 + \gamma^2 \tilde{\omega}^2 \beta^2} - 2 \mathrm{Im}(k)\right] \mathrm{d}r^\prime\right\},
\end{aligned}
\ee
where
\begin{align}
\label{eq:imk_cool}
\mathrm{Im}(k) & = \frac{(\gamma-1)\Omega\tilde{\omega}\beta}{2(\Omega^2 + \gamma\tilde{\omega}^2\beta^2)} \mathrm{Re}(k), \\
\label{eq:rek_cool}
\mathrm{Re}(k) & = \frac{|D|^{1/2}}{c_\mathrm{s,eff}},
\end{align}
are the imaginary and real parts of the radial wavenumber $k$, $\beta = \Omega t_\mathrm{c}$ is the dimensionless cooling time, and $r_0$ is an arbitrary reference radius. We have introduced the effective sound speed
\be
\label{eq:cs-eff}
c_\mathrm{s,eff} = \left(\frac{\Omega^2+\gamma^2\tilde{\omega}^2 \beta^2}{\Omega^2+\tilde{\omega}^2 \beta^2}\right)^{1/4} c_\mathrm{s,iso},
\ee
which varies from $c_\mathrm{s,iso}$ for $\beta = 0$ to $\gamma^{1/2}c_\mathrm{s,iso} = c_\mathrm{s,adi}$ as $\beta \rightarrow \infty$. Equation~(\ref{eq:fj_law_cool}) reduces to $F_J^m = $ constant in the adiabatic limit $\beta \rightarrow \infty$ or the globally isothermal limit $L_T \rightarrow \infty$ and $\gamma = 1$, and to $F_J^m \propto c_\mathrm{s}^2$ in the locally isothermal limit $\beta \rightarrow 0$. The accuracy with which the WKB analysis reproduces the true behavior of the AMF components is examined in Appendix \ref{sect:WKB-validate}.

Note that equation~(\ref{eq:fj_law_cool}) has explicit dependence on the azimuthal number $m$, unlike the equivalent expression for adiabatic (equation~(\ref{eq:fj_law_adi})) or locally isothermal disks (equation~(\ref{eq:fj_law_iso})). Therefore, the behavior of the total $F_J$, found by summing up the contributions for all azimuthal numbers, does not follow trivially from equation~(\ref{eq:fj_law_cool}). In general, the variation of $F_J$ depends on the relative magnitudes of all of the $F_J^m$ components at some reference radius.

\subsubsection{Inner Disk}

\begin{figure}
\begin{center}
\includegraphics[width=0.49\textwidth,clip]{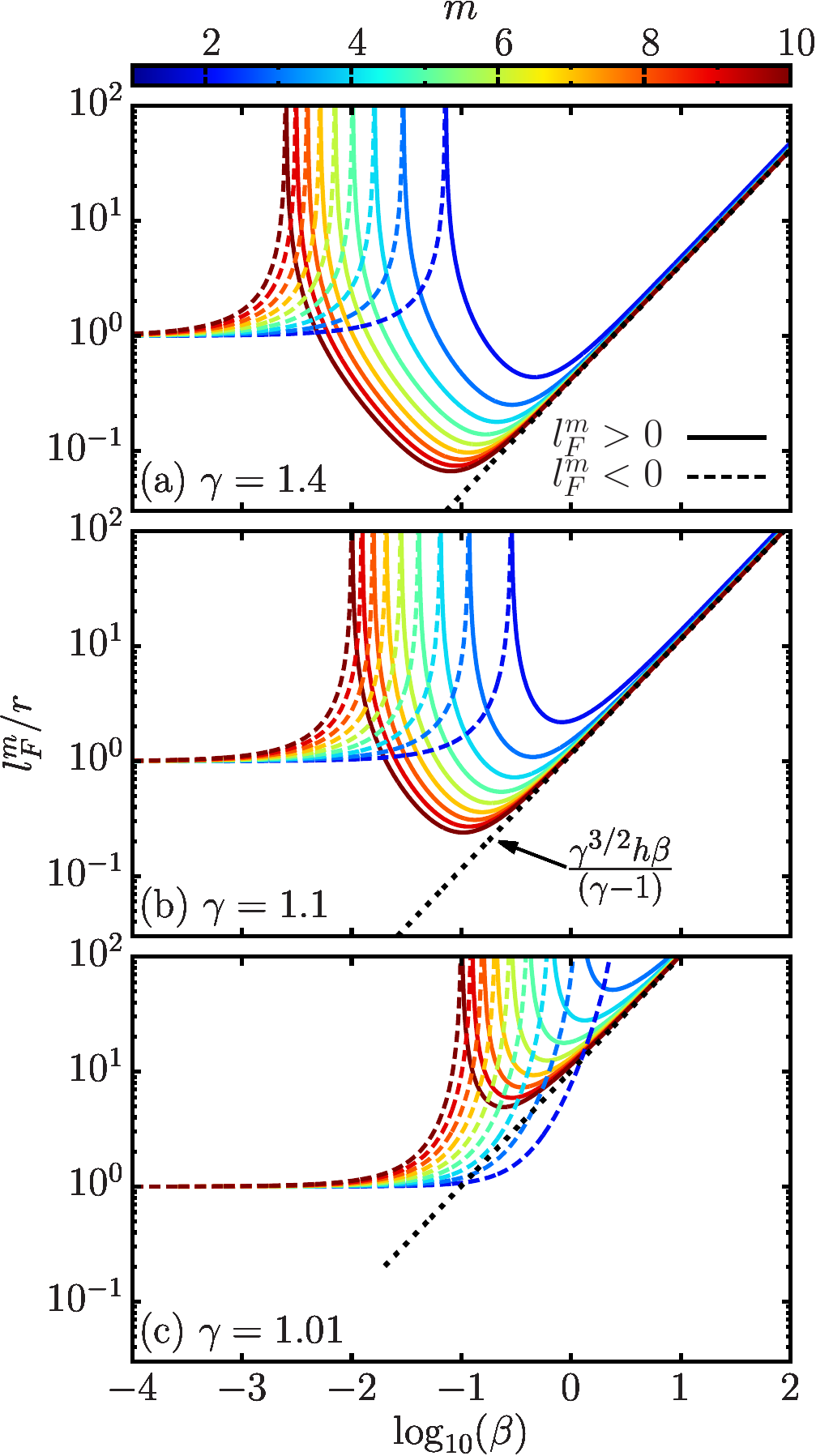}
\caption{Asymptotic behavior of $F_J^m$, the AMF of Fourier modes, for free waves in the WKB limit in the inner disk, where the wave pattern frequency is slow compared to the orbital frequency. The length scale $l_F^m$ (defined by equation~(\ref{eq:lfm}) and normalized by $r$) associated with the variation of $F_J^m$ is shown as a function of the dimensionless cooling timescale $\beta$ for different azimuthal mode numbers $m$ (different colored lines), and for different values of the adiabatic index $\gamma$ (different panels). The disk has a temperature power law index $q = 1$ and aspect ratio $h = 0.1$. Solid lines denote that $l_F^m$ is positive, so that $F_J^m$ decreases toward the center of the disk, and dashed lines denote that $l_F^m$ is negative, hence $F_J^m$ increases towards the center. The dotted line in each panel (labeled in (b)) indicates the asymptotic behavior for large $\beta$ given by equation (\ref{eq:lfm-betalarge}).}
\label{fig:amf_inner}
\end{center}
\end{figure}

In general, even the WKB equation~(\ref{eq:fj_law_cool}) must be evaluated numerically to determine the behavior of $F_J^m$. However, its asymptotic behavior in the inner disk can be expressed in a simple analytic form if we assume that $\beta$ is a constant, i.e., that the cooling timescale is a fixed fraction of the orbital period, as in our numerical calculations. A simple expression for the behavior of $F_J^m$ in the inner disk is of interest because of the complex wave phenomena that occur there (e.g., multiple spiral arms). The behavior of $F_J^m$ in the outer disk cannot be described by a simple expression for constant $\beta$, and so we leave detailed characterization of the AMF evolution in this region to the numerical results presented in Section~\ref{sect:amf-numerical}. 

At small radii in the inner disk, $\Omega \gg \omega_\mathrm{p}$ and hence $\tilde{\omega} \approx -m\Omega$. Making use of this approximation, the WKB equation~(\ref{eq:fj_law_cool}) reduces to
\be
\label{eq:fj_law_cool_inner}
F_J^m(r) = F_J^m(r_0) \left[\frac{c_\mathrm{s}(r)}{c_\mathrm{s}(r_0)}\right]^{2/(1+\gamma^2 m^2 \beta^2)} \exp\left[\xi g\left(\frac{r}{r_0}\right)\right],
\ee
where
\be
\xi = \frac{(\gamma-1)(m^2-1)^{1/2}m\beta}{h_0(1+\gamma m^2\beta^2)} \left(\frac{1+m^2\beta^2}{1+\gamma^2 m^2 \beta^2}\right)^{1/4},
\label{eq:xi}
\ee
with $h_0 = h(r_0)$, and
\be
g(x) = \begin{cases} 2(q-1)^{-1}\left[x^{(q-1)/2} - 1\right] & (q \neq 1), \\
\ln x & (q = 1).
\end{cases}
\ee
For the case $q = 1$, further simplification is possible:
\be
\label{eq:fj_law_cool_inner_q1}
F_J^m(r) = F_J^m(r_0) \left(\frac{r}{r_0}\right)^{\xi- 1/(1+\gamma^2 m^2 \beta^2)}.
\ee

There are two distinct contributions to the radial variation of the wave AMF for free waves in disks with cooling (equation~(\ref{eq:fj_law_cool})). The first is related to the disk temperature profile, described by the term in equation~(\ref{eq:fj_law_cool}) involving $L_T$. As typically the disk temperature varies on a global scale, i.e., $L_T \sim r$, this usually constitutes a slow variation. The second contribution formally describes a linear damping, as it is associated with an imaginary part of the radial wavenumber (equation~(\ref{eq:imk_cool})). Damping occurs because thermal relaxation counteracts the adiabatic heating/cooling associated with compression/expansion of a fluid element, reducing the restoring action of the pressure for the wave.

We characterize this damping by examining the asymptotic behavior of the WKB AMF $F_J^m$ in the inner disk, as $r \to 0$. Using equation~(\ref{eq:fj_law_cool_inner}), we define the characteristic length scale of the radial variation of the AMF,
\be
\label{eq:lfm}
l_F^m = \left(\frac{\mathrm{d}\ln F_J^m}{\mathrm{d}r}\right)^{-1}.
\ee
Note that the sign of $l_F^m$ is important, as it indicates whether the magnitude of $F_J^m$ increases or decreases with radius. For inward traveling waves in the inner disk, a positive $l_F^m$ means that $F_J^m$ decreases as the wave propagates toward the inner disk, and a negative $l_F^m$ means that it grows as the wave propagates inward.

We first focus on the case of a disk with a $q = 1$ temperature profile, which has a constant aspect ratio $h_0=h=$ const in equation (\ref{eq:xi}). In this case, $F_m$ has a power law dependence on $r$ in the inner disk (equation~(\ref{eq:fj_law_cool_inner_q1})). Therefore, AMF varies with a length scale proportional to $r$, i.e., $l_F^m/r$ is independent of radius. Specifically, from equation~(\ref{eq:fj_law_cool_inner_q1}), we have
\be
\label{eq:lfm-q1}
\frac{l_F^m}{r} = \left(\xi - \frac{1}{1 + \gamma^2 m^2 \beta^2}\right)^{-1}.
\ee
Because of this simplification, a disk model with $q = 1$ serves as a convenient case for analysis. Fig.~\ref{fig:amf_inner} shows $l_F^m/r$ as a function of $\beta$ for waves with different azimuthal numbers in the inner disk for a disk with $q = 1$ and aspect ratio $h = 0.1$. The azimuthal numbers shown are $m = 2$ -- $10$, which carry most of the AMF for planet-excited waves in a disk with this aspect ratio (note that $m = 1$ waves are evanescent in the inner disk).

In Fig.~\ref{fig:amf_inner}(a), we consider the case with $\gamma = 1.4$, appropriate for protoplanetary disks. We highlight several key features of the behavior of the $F_J^m$ illustrated by this figure. First, for this $\gamma$ there is always a range of values of $\beta$ for which $l_F^m/r$ is positive and $\lesssim 1$; for the plotted values of $m$ this range falls into an interval $10^{-2} \lesssim \beta \lesssim 1$, with higher $m$ modes reaching lower $l_F^m/r$. This means that $F_J^m$ decreases rapidly towards the inner disk, i.e., waves are {\it damped}, with a damping length significantly shorter than $r$, for all values of $m$. 

Second, for large values of $\beta$ ($\gtrsim 1$), $l_F^m$ is positive and increases with $\beta$. For $\beta \gtrsim 10$, $l_F^m/r > 1$, corresponding to weak damping, and for $\beta \gtrsim 100$, $l_F^m/r \gtrsim 10$, so that the waves are effectively undamped and the adiabatic limit is recovered. Note also that $l_F^m$ is nearly independent of $m$ for values of $\beta$ in this range. 

Finally, for small values of $\beta$ (depending on $m$ but $\lesssim 10^{-1}$), $l_F^m$ is negative, so that $F_J^m$ {\it grows} as the wave propagates towards the inner disk. For $\beta \approx 10^{-4}$, $l_F^m \approx -1$ (i.e., $F_J^m \propto 1/r$) for all values of $m$ that we consider, so that the locally isothermal limit is recovered. Note that even for $\beta$ as small as $10^{-3}$, $l_F^m$ still shows significant variation with $m$, meaning that the locally isothermal limit is not valid.

The details of this behavior can be understood by examining equation~(\ref{eq:lfm-q1}). We find that the minimum positive $l_F^m$, i.e., the shortest possible damping length, occurs when $\beta = \beta_\mathrm{crit}$, where
\be
\label{eq:beta_crit}
\beta_\mathrm{crit} \approx \frac{1}{\gamma^{1/2}m}.
\ee
Note that, aside from the order unity factor $\gamma^{1/2}$, $\beta_\mathrm{crit}$ can be interpreted as an approximate equality of the cooling timescale $t_\mathrm{c}$ and $1/(m\Omega)$, the time for a fluid element to cross through one wavelength of the $m$-fold perturbation (in the inner disk).

When $\beta = \beta_\mathrm{crit}$, the damping coefficient $\xi$ in equation~(\ref{eq:fj_law_cool_inner}) reaches 
\be
\xi_\mathrm{max} = \frac{(\gamma-1)(m^2-1)^{1/2}}{2\gamma^{3/4}h},
\ee
and the shortest possible damping length is
\be
\label{eq:ldmin}
l_\mathrm{d,min}^m \approx \frac{r}{\xi_\mathrm{max}} = \frac{2 \gamma^{3/4} H}{(\gamma-1)(m^2-1)^{1/2}}.
\ee
Note that $l_\mathrm{d,min}^m\propto m^{-1}$ for large $m$, just as observed in Fig. \ref{fig:amf_inner}. For planet-driven waves in a disk with aspect ratio $h_\mathrm{p}$, the dominant mode has $m \approx m_* \approx (2h_\mathrm{p})^{-1}$ \citep{OL02}. For this mode, $\beta_\mathrm{crit} \approx 2 h_\mathrm{p}/\gamma^{1/2}$. For $\gamma = 7/5$ and $h = 0.1$, the minimum damping length for the $m_*$ mode is $l_\mathrm{d,min} \approx 1.3 H_\mathrm{p}$, i.e., very short. 

Using equation~(\ref{eq:lfm-q1}), we also find that the asymptotic behavior for large values of $\beta$ is
\be
\label{eq:lfm-betalarge}
\left(\frac{l_F^m}{r}\right)_{\beta \gg \beta_\mathrm{crit}} \approx \frac{\gamma^{3/2}h\beta}{\gamma-1},
\ee
which is independent of the value of $m$. This behavior is in perfect agreement with the results shown in Fig.~\ref{fig:amf_inner}.

Additionally, $l_F^m$ becomes negative for $\beta \lesssim \beta_\pm$, where
\be
\label{eq:beta-pm}
\beta_\pm \approx \frac{h}{(\gamma-1)m^2}.
\ee
For planet-driven waves with $m = m_*$ carrying most of the angular momentum, 
\ba
\beta_\pm(m_*) \approx \frac{4}{\gamma-1}h_\mathrm{p}^3,
\label{eq:beta-to-iso}
\ea
which is about $10h_\mathrm{p}^3$ for $\gamma = 7/5$.

This value of the cooling rate has a very important meaning: it sets an {\it upper limit} on the value of $\beta$ below which the AMF behavior begins to converge to the locally isothermal regime. One can see from Fig.~\ref{fig:amf_inner} that $\beta$ has to be at least an order of magnitude smaller than the value $\beta_\pm(m_*)$, at which $l_F^m$ changes sign, for $l_F^m/r$ to finally converge to $-1$, which signals the ultimate transition to the locally isothermal behavior. Thus, it is natural to expect that the locally isothermal behavior can be reproduced in disks with cooling only when $\beta \lesssim 0.1\beta_\pm(m_*)\sim h_\mathrm{p}^3$. This point is further discussed in Section~\ref{sect:other}.

The case of a smaller value for the adiabatic index, $\gamma = 1.1$ (chosen only to explore the dependence on $\gamma$), is shown in Fig.~\ref{fig:amf_inner}(b). We see that the behavior of $F_J^m$ is qualitatively very similar to the case with $\gamma = 1.4$. In particular, there are regions of strong wave damping for $\beta$ in the range $10^{-2} \lesssim \beta \lesssim 1$. In agreement with equation~(\ref{eq:ldmin}), the damping is weaker, i.e., $l_F^m/r$ is larger by a factor of a few as compared to the case with $\gamma = 1.4$, but nonetheless $l_F^m/r \ll 1$ for these values of $\beta$. In this case, the locally isothermal limit is reached for $\beta \lesssim 3\times 10^{-3}$---larger than the limiting value for $\gamma = 1.4$, in agreement with equation~(\ref{eq:beta-pm}). 

Finally, for $\gamma$ very close to unity, $\gamma = 1.01$ (see Fig.~\ref{fig:amf_inner}(c)), the damping is very weak for all values of $\beta$. The behavior of $F_J^m$ transitions almost monotonically from the locally isothermal limit to the adiabatic limit in this case. Note that for this $\gamma$, the locally isothermal limit is valid for $\beta \lesssim 0.03$. This is much larger than the limiting value for $\gamma = 1.4$, again due to the $(\gamma-1)^{-1}$ dependence of $\beta_\pm$ (equation~(\ref{eq:beta-pm})), but it is still $\ll 1$.

The preceding analysis has focused on the case of a $q = 1$ temperature profile. For $q \neq 1$, for which $l_F^m/r$ varies with $r$, the behavior is qualitatively similar to the $q = 1$ case shown in Fig.~\ref{fig:amf_inner}, but quantitatively modified in two key ways (as described by equations~(\ref{eq:fj_law_cool_inner}) and (\ref{eq:lfm})). First, since the minimum damping length $l_\mathrm{d,min}$ is proportional to the local scale height $H(r)$ (see equation~(\ref{eq:ldmin}), which is valid for general values of $q$), $l_\mathrm{d,min}/r$ is proportional to the local aspect ratio $h(r)$. This is independent of $r$ for $q = 1$ but $\propto r^{(1-q)/2}$ in general. Second, for small values of $\beta$, in the nearly locally isothermal limit, in general $l_F^m/r \rightarrow -1/q$ for all values of $m$. The limiting value of $\beta$ at which this transition occurs is still given approximately by equation~(\ref{eq:beta-pm}), but modified by a factor of $q$. This is a relatively minor effect, since $q$ lies within a fairly narrow range of values (i.e., varying by tens of per cent) for most physically reasonable disk models.

\subsection{Numerical Results}
\label{sect:amf-numerical}

\begin{figure*}
\begin{center}
\includegraphics[width=0.99\textwidth,clip]{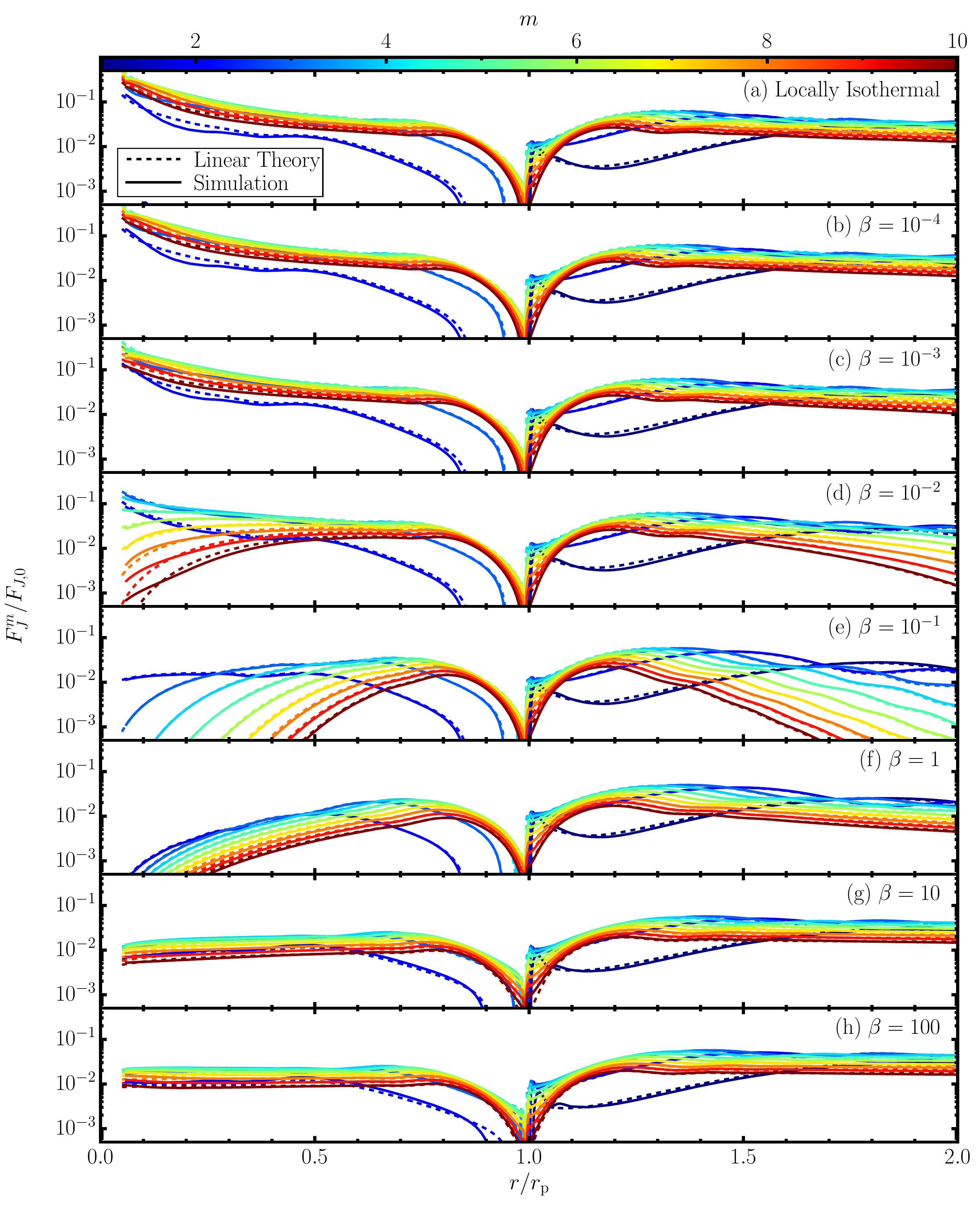}
\caption{Profiles of the Fourier components of the wave AMF $F_J^m$, in terms of the characteristic scale $F_{J,0}$ (equation~(\ref{eq:fj0})), for a disk with adiabatic index $\gamma = 7/5$, aspect ratio $h_\mathrm{p} = 0.1$, temperature power law index $q = 1$, and different thermodynamics: locally isothermal and cooling with different values of $\beta$. Solid lines of different colors, corresponding to different azimuthal numbers $m$, are the results of numerical simulations with a $0.01 M_\mathrm{th}$ planet at $20$ orbits. The corresponding dashed lines are the $F_J^m$ components for planet-excited waves computed from linear theory.}
\label{fig:amf_fourier}
\end{center}
\end{figure*}

\begin{figure*}
\begin{center}
\includegraphics[width=0.99\textwidth,clip]{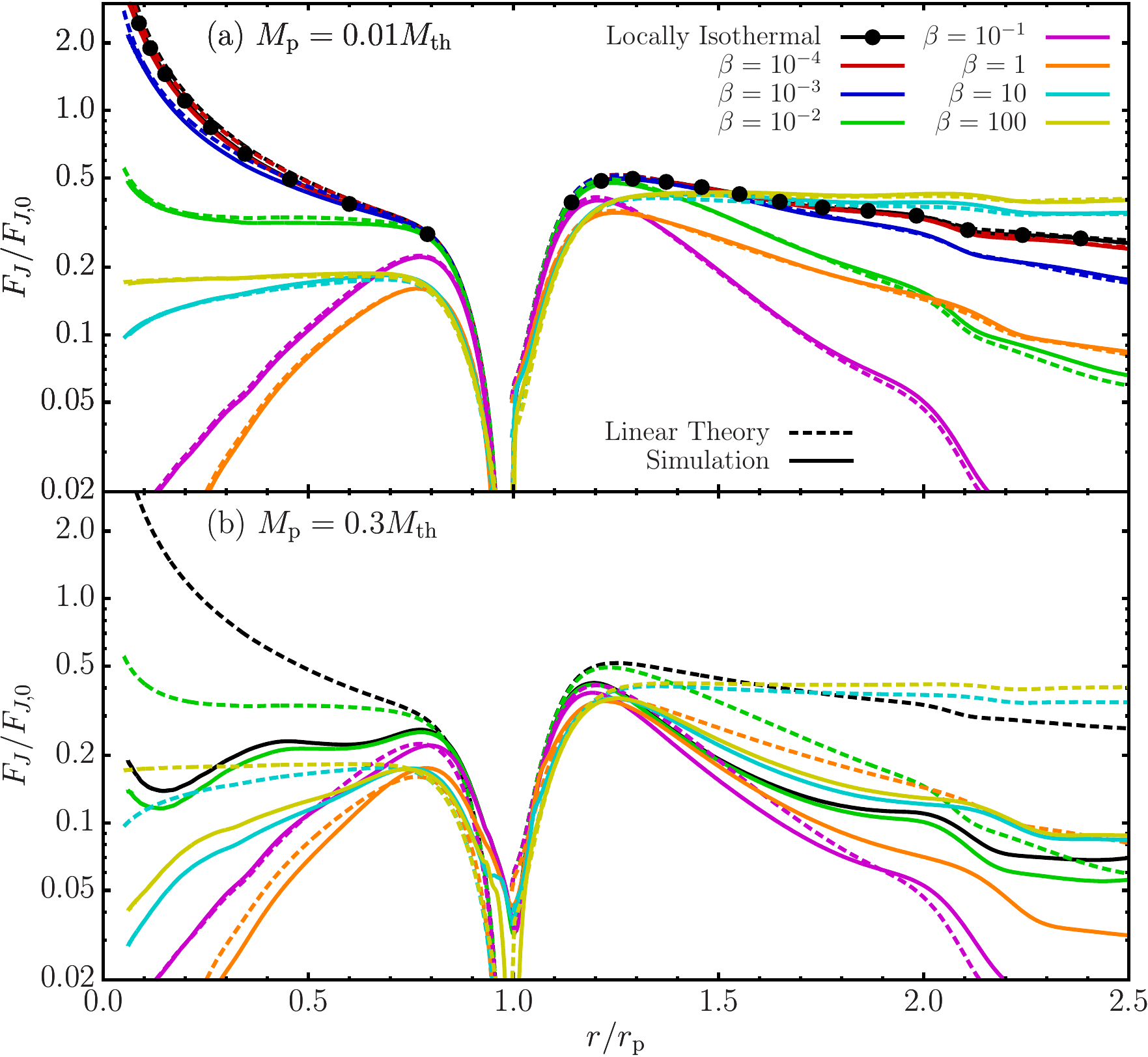}
\caption{The total AMF $F_J$ for planet-driven waves in disks with different thermodynamics (the same cases as in Fig.~\ref{fig:amf_fourier}). The dashed lines, which are the same in both panels, are computed using linear theory. The solid lines are the results of numerical simulations, with a $0.01 M_\mathrm{th}$ planet in (a), and a $0.3 M_\mathrm{th}$ planet in (b), taken at $20$ orbits. The filled points in (a) highlight the curve for the locally isothermal simulation, in order to help distinguish it from the other similar curves for small values of $\beta$.}
\label{fig:amf_total}
\end{center}
\end{figure*}

\begin{figure}
\begin{center}
\includegraphics[width=0.49\textwidth,clip]{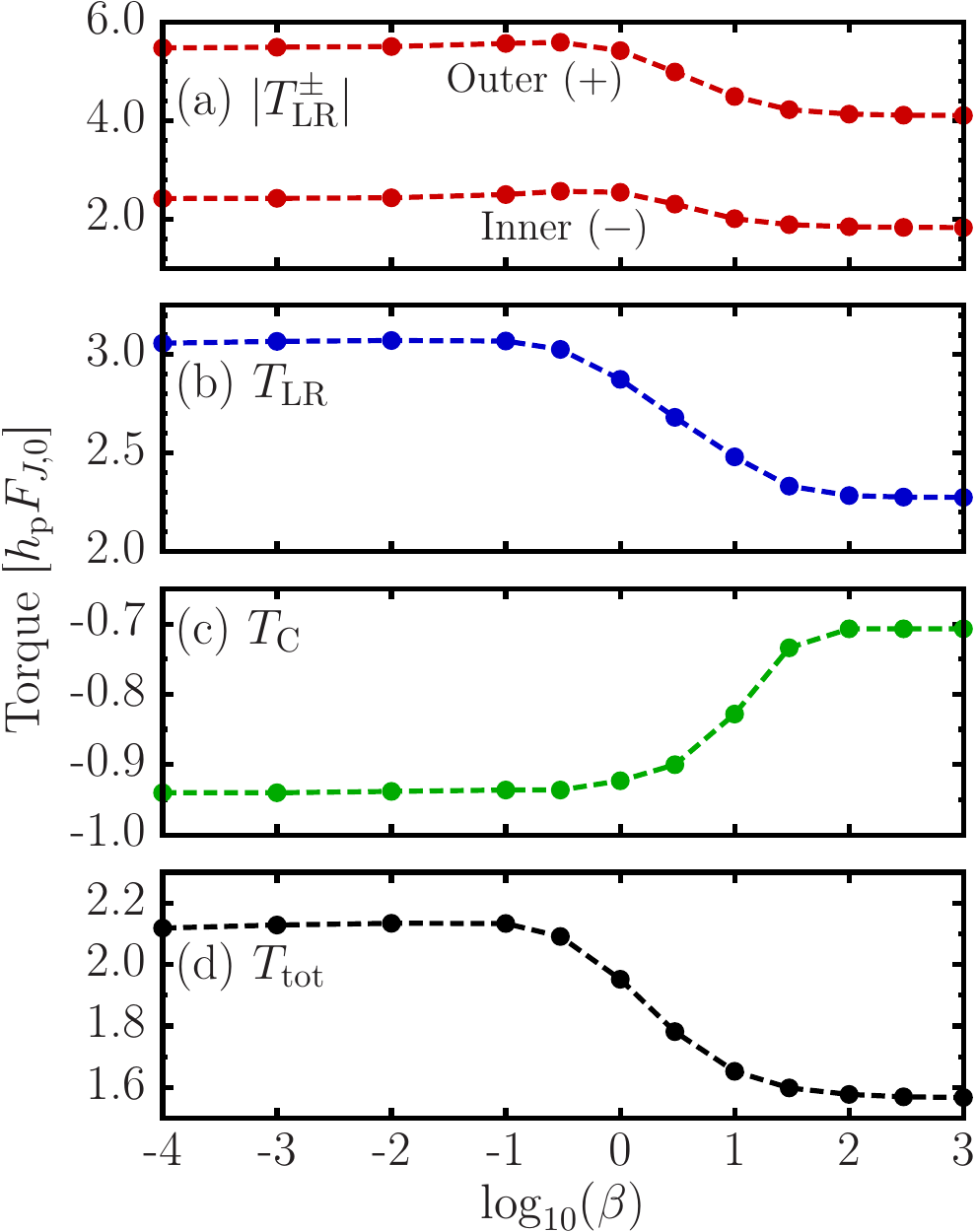}
\caption{Different components of the torque exerted by the planet on the disk, computed using linear theory, as a function of $\beta = \Omega t_\mathrm{c}$: (a) magnitude of the one-sided Lindblad torques $T_\mathrm{LR}^\pm$ (note that $T_\mathrm{LR}^- < 0$), (b) differential Lindblad torque $T_\mathrm{LR}$, (c) corotation torque $T_\mathrm{C}$, and (d) total torque $T_\mathrm{tot}$. See Section~\ref{sect:torque} for details. The disk has an aspect ratio $h_\mathrm{p} = 0.1$, temperature and surface density power law indices $q = 1$ and $p = 1$, and adiabatic index $\gamma = 7/5$.}
\label{fig:torque}
\end{center}
\end{figure}

Profiles of the Fourier AMF $F_J^m$ for planet-excited waves are shown in Fig.~\ref{fig:amf_fourier}, for disks with varied thermodynamics: a locally isothermal disk, and disks with cooling characterized by different values of $\beta$ ranging from $10^{-4}$ to $10^2$. The AMF is expressed in terms of the characteristic scale 
\be
\label{eq:fj0}
F_{J,0} = \left(\frac{M_\mathrm{p}}{M_*}\right)^2 h_\mathrm{p}^{-3} \Sigma_\mathrm{p} r_\mathrm{p}^4 \Omega_\mathrm{p}^2,
\ee
associated with the total angular momentum transfer at Lindblad resonances \citep{GT80,Ward1997}. In each panel, the results of numerical simulations (Section~\ref{sect:sims}) for $M_{\rm p} \ll M_{\rm th}$, when the perturbation is well in the linear regime, are shown as solid lines, with different colors representing different azimuthal mode numbers. Dashed lines show the results of linear theory obtained by numerically solving the master equation (\ref{eq:master}) with the isothermal (equations (\ref{eq:c1_iso})--(\ref{eq:LT})) or cooling (equations (\ref{eq:c1_cool})--(\ref{eq:Dc})) inputs and fully accounting for the forcing by the planetary potential (see Section~\ref{sect:lin-calc}). Clearly, there is excellent agreement between the theoretical and numerical results, again validating our theoretical analysis. Note that the linear solutions self-consistently capture the behavior of $F_J^m$ in the vicinity of the planet where wave excitation occurs. Beyond this region, the WKB approximation (see Section~\ref{sect:amf-cool}) provides an excellent description of the evolution of $F_J^m$, as we demonstrate in Appendix~\ref{sect:WKB-validate}. Profiles of the total AMF (resulting from the sum of all Fourier modes) are shown in Fig.~\ref{fig:amf_total}(a) (for the same cases as shown in Fig.~\ref{fig:amf_fourier}). Here the solid and dashed lines again represent the results of numerical simulations and linear calculations, respectively, which show close agreement with one another.

We first examine the AMF in the inner disk. The AMF behavior for a locally isothermal disk (Fig.~\ref{fig:amf_fourier}(a)), for which $F_J^m \propto r^{-1}$ (for $q = 1$) for all $m$, is reproduced very closely for the case $\beta = 10^{-4}$ (Fig.~\ref{fig:amf_fourier}(b)). The case with $\beta = 10^{-3}$ (Fig.~\ref{fig:amf_fourier}(c)) is qualitatively very similar, although there are slight differences; the different $F_J^m$ components, and hence the total $F_J$ (see Fig.~\ref{fig:amf_total}(a)), do not increase toward the inner disk quite as steeply.

For $\beta = 10^{-2}$ (Fig.~\ref{fig:amf_fourier}(d)), the results are substantially different. In this case, the damping rates of the different $F_J^m$ are very different from one another. For modes with $m \lesssim 5$, $F_J^m$ increases toward the inner disk, while for $m \gtrsim 5$ it decreases as a result of cooling-related damping. As a result, the total AMF (see Fig.~\ref{fig:amf_total}(a)), is approximately constant in the inner disk. The approximate behavior of the total AMF can be qualitatively deduced from the behavior of $F_J^{m_*}$, where $m_* = 5$ (for $h_\mathrm{p} = 0.1$) is the dominant mode for planet-excited waves. For $\beta = 10^{-1}$ (Fig.~\ref{fig:amf_fourier}(e)), all of the $F_J^m$ decrease toward the inner disk ($l_F^m>0$), at a rate that increases with $m$. The resulting total AMF therefore decreases rapidly toward the inner disk.

For $\beta = 1$ (Fig.~\ref{fig:amf_fourier}(f)), the modes all have similar damping rates. For $\beta = 10$ (Fig.~\ref{fig:amf_fourier}(g)) all of the $F_J^m$ behave essentially the same, exhibiting only a weak decay toward the inner disk. Finally, for $\beta = 100$ (Fig.~\ref{fig:amf_fourier}(h)), the AMF of all the modes is approximately constant, and hence so is the total AMF, effectively reproducing the behavior expected in a purely adiabatic disk. 

In the outer disk, the behavior of the AMF is somewhat simpler. This is because the effects of cooling can only act to make $F_J^m$ can decrease with $r$ in the outer disk (it would decrease even in the locally isothermal case). This is a result of either the AMF following or nearly following the global temperature gradient, e.g., for $\beta = 10^{-4} - 10^{-3}$, or due to linear damping, e.g., for $\beta = 10^{-2}$ -- $10$. For intermediate values of $\beta$, the different modes have different damping rates in the outer disk, although the differences are less pronounced than in the inner disk.

Taken together, the results for the AMF illustrate the fact that the cooling time must be extremely short relative to the orbital timescale, with $\beta \lesssim 10^{-3}$, for the locally isothermal approximation to provide an accurate description of density waves in the linear regime. The AMF behavior for purely adiabatic disks is reproduced for $\beta \gtrsim 10$. There is therefore a very wide range of cooling times, with $\beta$ spanning about four orders of magnitude, for which neither the adiabatic nor locally isothermal approximations provide a good description of the density wave dynamics.

\subsubsection{Torque}
\label{sect:torque}

Our analysis has so far been primarily concerned with the radial variation of the wave AMF due to cooling that occurs outside the wave excitation zone. However, as seen in Fig.~\ref{fig:amf_total}, the ``initial amplitude'' of $F_J$ (i.e., its value a few scale heights away from the planet) varies with the cooling timescale. This is indicative of a variation of the torque exerted on the disk with $\beta$. As this torque is eventually responsible for planet migration, we would like to quantify how it is affected by cooling. The net torque on the disk is
\be
T_\mathrm{tot} = \int_{r_\mathrm{in}}^{r_\mathrm{out}} \frac{\mathrm{d}T}{\mathrm{d}r} \mathrm{d}r,
\ee
where $\mathrm{d}T/\mathrm{d}r$ (equation~(\ref{eq:dtdr})) is the torque density. The orbital evolution of the planet is described by $\dot{L}_\mathrm{p} = -T_\mathrm{tot}$, where $L_\mathrm{p} = M_\mathrm{p}(GM_* r_\mathrm{p})^{1/2}$ is the angular momentum of the planet.

There are three different contributions to the net torque. These are the one-sided Lindblad torques $T_\mathrm{LR}^\pm$ (where $+$ and $-$ refer to the outer and inner disk), resulting from wave excitation at Lindblad resonances, and the corotation torque $T_\mathrm{C}$, associated with angular momentum transfer at the corotation radius $r_\mathrm{C}$, where $\Omega(r_\mathrm{C}) = \Omega_\mathrm{p}$ (e.g., \citealt{GT79}). These torques are computed using profiles of the total (i.e., summed over $m$) AMF $F_J$ and torque density $\mathrm{d}T/\mathrm{d}r$ computed from linear theory. We forgo an analysis of the torques in our numerical simulations, as the corotation torque is subject to oscillations and saturation (e.g., \citealt{Paardekooper2008}) on timescales much longer than we have simulated.

In adiabatic disks, the one-sided Lindblad torques are identified as the asymptotic values of $F_J$ at large distances from the planet. This is a consequence of the conservation of AMF for free waves---in adiabatic disks the wave AMF changes only as a result of excitation by the planetary potential, described by $\mathrm{d}T/\mathrm{d}r$, except in the immediate vicinity of the corotation resonance, where it experiences a discontinuous jump. As we have shown, cooling results in additional radial variation of $F_J$, described by the difference between $\mathrm{d}F_J/\mathrm{d}r$ and $\mathrm{d}T/\mathrm{d}r$. Therefore, the one-sided Lindblad torques can be computed via the following formula:
\be
\label{eq:tlr}
T_\mathrm{LR}^\pm = \pm F_J(r_\mathrm{out/in}) \mp \int_{r_\mathrm{C}^\pm}^{r_\mathrm{out/in}} \left(\frac{\mathrm{d}F_J}{\mathrm{d}r} - \frac{\mathrm{d}T}{\mathrm{d}r}\right) \mathrm{d}r.
\ee
The second term in equation~(\ref{eq:tlr}) compensates for the decay (or growth) of AMF due to cooling that occurs between just outside the corotation radius and the disk edge. The lower limit of integration in this term is $r_\mathrm{C}^\pm = r_\mathrm{C} \pm \epsilon$, with $0 < \epsilon \lesssim 0.005 r_\mathrm{p}$, chosen to cut out the jump in $F_J$ at $r_\mathrm{C}$ due to the corotation torque (which occurs over a finite radial distance in our numerical results). For $\beta \rightarrow \infty$, the integrand of the second term in equation~(\ref{eq:tlr}) vanishes, and we recover the usual definition of the one-sided Lindblad torque for adiabatic disks.

The total torque due to both the outer and inner Lindblad resonances---the so-called differential Lindblad torque---is then
\be
T_\mathrm{LR} = T_\mathrm{LR}^+ + T_\mathrm{LR}^-,
\ee
and finally the corotation torque is given by
\be
T_\mathrm{C} = T_\mathrm{tot} - T_\mathrm{LR}.
\ee

The results of this decomposition are shown in Fig.~\ref{fig:torque}. Results for several additional values of $\beta$ not shown in Figs.~\ref{fig:amf_fourier}--\ref{fig:amf_total} are displayed in order to fully illustrate the behavior of the torques as a function of $\beta$. All of the components of the torque shown in Fig.~\ref{fig:torque} exhibit approximately the same dependence on $\beta$. Each one is constant for $\beta \lesssim 10^{-1}$, undergoes a moderate decrease as $\beta$ is increased to $\approx 100$, and is again constant for $\beta \gtrsim 100$. In the locally isothermal limit ($\beta \rightarrow 0$), the torques are $30$ -- $35\%$ larger than in the adiabatic limit ($\beta \rightarrow \infty$). This is related to the variation of the effective sound speed $c_\mathrm{s,eff}$ (see equation~(\ref{eq:cs-eff})) from $c_\mathrm{s,iso}$ to $c_\mathrm{s,adi}$ as $\beta$ is increased. The smaller $c_\mathrm{s,eff}$ for small values of $\beta$ results in a stronger response of the disk to the planetary potential, and therefore a larger torque. Note that the total torque on the disk (Fig.~\ref{fig:torque}d) is always positive, resulting in inward migration of the planet.

Analytic studies of wave excitation at Lindblad resonances \citep{GT80,Ward1997} find the characteristic scale of the one-sided Lindblad torques $T_\mathrm{LR}^\pm$ to be $F_{J,0}$ (equation~(\ref{eq:fj0})), while the characteristic scale of the differential Lindblad torque $T_\mathrm{LR}$ should be smaller by a factor $h_\mathrm{p}$. Hence, $T_\mathrm{LR}^\pm \propto c_\mathrm{s}^{-3}$ and $T_\mathrm{LR} \propto c_\mathrm{s}^{-2}$. However, in practice $T_\mathrm{LR}^\pm$ and $T_\mathrm{LR}$ are found to be similar in magnitude when $h_\mathrm{p} \approx 0.1$  \citep{Ward1997,Papaloizou2007} as we have considered in this work. As such, all of the torques in Fig.~\ref{fig:torque} are shown in terms of the differential Lindblad torque scale $ h_\mathrm{p} F_{J,0}$.

The aforementioned scalings suggest that $T_\mathrm{LR}^\pm(\beta \rightarrow 0)/T_\mathrm{LR}^\pm(\beta \rightarrow \infty) \approx \gamma^{3/2}$ and $T_\mathrm{LR}(\beta \rightarrow 0)/T_\mathrm{LR}(\beta \rightarrow \infty) \approx \gamma$. In our calculations, $T_\mathrm{LR}$ obeys the expected scaling, whereas $T_\mathrm{LR}^\pm$ exhibits a somewhat weaker variation with $\beta$ than expected. This may be related to the the relatively large value of $h_\mathrm{p}$ used, or the softening of the planetary gravitational potential in our calculations, which is not included in many analytic studies. We note that our value for $T_\mathrm{LR}$ in the adiabatic limit is in good agreement (within a few percent) with the adiabatic Lindblad torque formula given by \citet{Paardekooper2010}, which is based on numerical solutions of the linear perturbation equations (including softening).

We conclude that cooling has only a a modest effect (tens of percent) on the magnitude of the linear torque associated with planet migration for the disk model we have considered. However, more dramatic effects associated with cooling might be possible. In particular, the corotation torque, which is prone to becoming nonlinear even for low-mass planets, is sensitive to the entropy gradient of the disk, and its behavior has previously been shown to be sensitive to cooling. Reversal of the direction of migration is possible when these effects are taken into account \citep{Paardekooper2008}. A thorough exploration of this topic, requiring an exploration of different disk profiles and a full consideration of nonlinear effects, is beyond the scope of this work.

\section{Massive Planets and Disk Evolution}
\label{sect:massive}

Simulations shown in the previous section explored the linear regime of the planet-disk coupling. We now explore numerically the role of cooling on the planet-disk interaction in the presence of the nonlinear effects. It is well known that density waves launched by massive planets in adiabatic disks undergo rapid nonlinear evolution resulting in their shocking and dissipation \citep{GR01,R02,Dong2011b}. Our goal will be to explore the interplay between the wave evolution due to cooling (which is a linear phenomenon) and due to the nonlinear effects. Another goal is to examine the effect of the different levels of cooling on the long-term evolution of the disk---variation of its surface density caused by the deposition of the angular momentum of the density wake.

We consider a planet mass which is a moderate fraction of the thermal mass, $M_\mathrm{p} = 0.3 M_\mathrm{th}$. The setup for these simulations is the same as described in Section~\ref{sect:sims}, except that we reduce the numerical resolution by a factor of two, to $N_r \times N_\phi = 1502 \times 2048$, in order to facilitate a longer evolution timescale of $500 t_\mathrm{p}$. This is long enough for the surface density profile of the disk to undergo significant evolution due to the planet-driven waves. We make use of profiles of the wave AMF as a diagnostic of the wave-driven evolution of the disk (e.g., \citealt{Miranda-ALMA}). As before, we consider both locally isothermal disks and disks with cooling with several values of $\beta$, which are representative of the variety of different behaviors of $F_J$ in the linear regime studied before (\S \ref{sect:amf-numerical}).

\subsection{AMF Profiles}
\label{sect:AMF_massive}

Fig.~\ref{fig:amf_total}(b) shows the profiles of the total AMF at $20$ orbits (solid lines), along with the AMF profiles computed using linear theory (dashed lines). For the locally isothermal case, $F_J$ begins to deviate from the linear prediction at about one or two scale heights from the planet, which is consistent with the theoretical shocking distance $l_\mathrm{sh} \approx (M_\mathrm{p}/M_\mathrm{th})^{-2/5} \approx 1.6 H_p$ \citep{GR01}. Beyond this distance, the actual $F_J$ is always smaller than the linear $F_J$, as a result of nonlinear dissipation. In most other cases (i.e., for different values of $\beta$), $F_J$ also begins to drop below the linear prediction at a distance $\approx l_\mathrm{sh}$ from the planet.

Two particular cases shown in Fig.~\ref{fig:amf_total}(b), locally isothermal and $\beta = 10^{-2}$, have very similar AMF profiles, despite the fact that the corresponding linear AMF profiles are quite different. Given that the linear AMF for the locally isothermal case is always larger than it is for the case with $\beta = 10^{-2}$, this indicates that in the locally isothermal case, the waves experience more nonlinear dissipation, resulting in the actual $F_J$ being comparable between these two cases almost everywhere in the disk. This similarity is probably coincidental, resulting from the particular strength of the nonlinear dissipation for our chosen planet mass, and would not occur for a different mass. Note however that, regardless of these details, the fact that the AMF profile is nearly the same for these two cases indicates that the resulting wave-driven disk evolution should also be nearly the same. 

The most important feature of Fig.~\ref{fig:amf_total}(b) is the fact that for the case with $\beta = 10^{-1}$, the AMF profile is not substantially different from the corresponding linear AMF profile. This is also true for case with $\beta = 1$, although mostly in the inner disk---there is some deviation in the outer disk. This indicates that nonlinear dissipation plays a much smaller role in the evolution of the density waves in these cases, i.e., for values of $\beta$ in this range. Instead, evolution of the wave AMF is mostly controlled by the linear damping associated with cooling. Evidently this damping is so strong that it is the dominant source of dissipation even for a fairly massive planet such as the one considered here ($0.3 M_\mathrm{th}$).

The fact that nonlinear dissipation is subdominant to linear dissipation for $\beta \approx 10^{-1}$ -- $1$ suggests that wave-driven evolution of the disk operates differently for disks with dimensionless cooling timescales in this range. We may therefore expect one of two different types of disk evolution, one associated with nonlinear wave dissipation (for very long or very short cooling timescales), and one dominated by linear wave dissipation (in the aforementioned range of $\beta$). The exact separation between these two regimes in terms of the cooling timescale should be a function of $M_{\rm p}/M_{\rm th}$.

\begin{figure}
\begin{center}
\includegraphics[width=0.49\textwidth,clip]{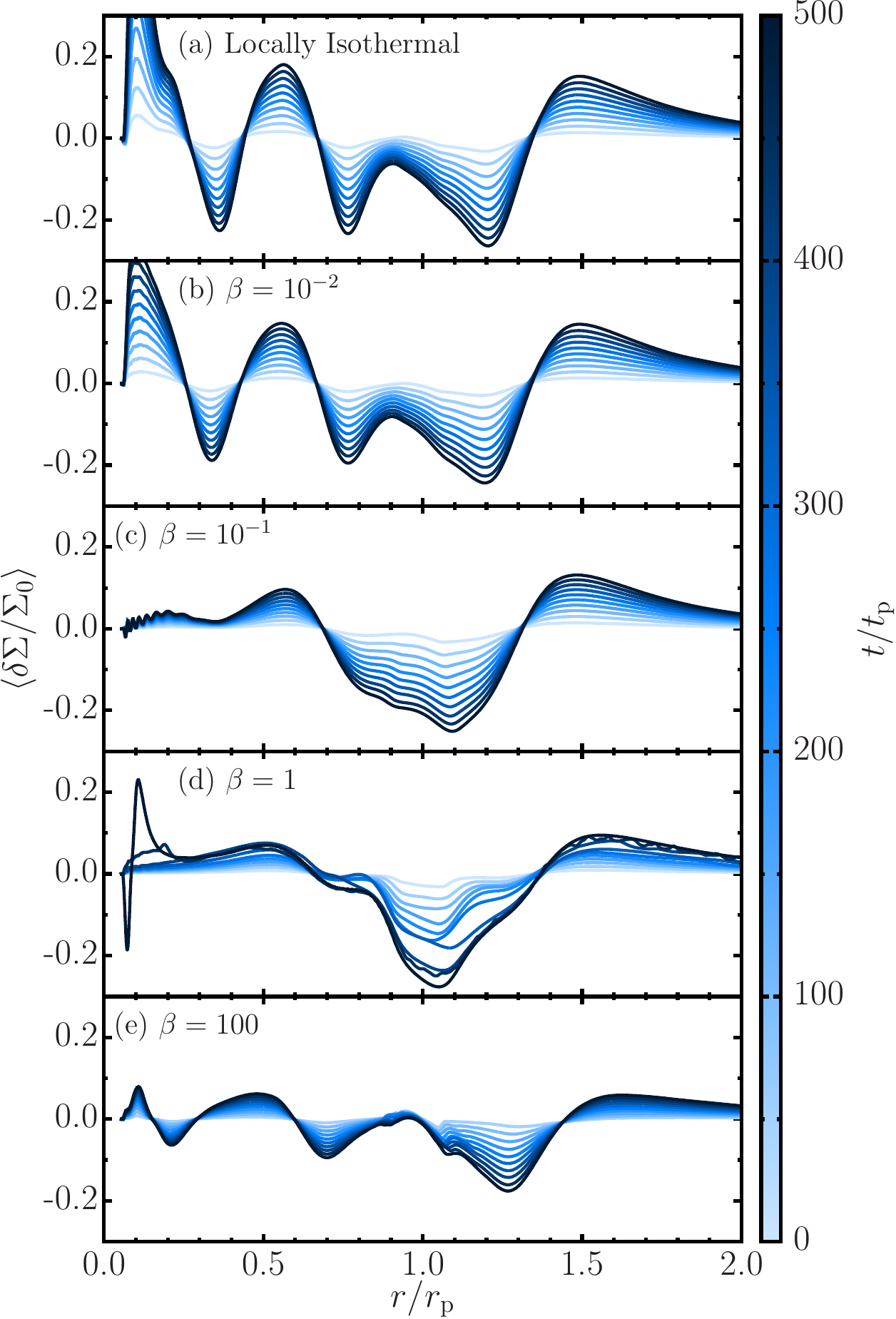}
\caption{Evolution of the disk surface density for $M_\mathrm{p} = 0.3 M_\mathrm{th}$ in disks with different thermodynamics. The azimuthally averaged surface density perturbation $\delta\Sigma$, relative to the the initial surface density profile $\Sigma_0$, is shown every $50$ orbits for $500$ orbits.}
\label{fig:evolution}
\end{center}
\end{figure}

\subsection{Surface Density Evolution}

Fig.~\ref{fig:evolution} shows the evolution of the disk surface density profile for the different $\beta$ cases shown in Fig~\ref{fig:amf_total}(b). There is a distinct dichotomy in the resultant disk structures. For sufficiently short and sufficiently long cooling times (Fig.~\ref{fig:evolution}(a),(b),(e)), for which wave dissipation happens primarily through shocks, the disk exhibits multiple gap and ring structures. For these cases, three gaps (surface density minima) and four rings (surface density maxima) are formed. Note that the surface density perturbation in the case of $\beta=10^{-2}$ is almost the same as in the isothermal case, which is to be expected based on Fig.~\ref{fig:amf_total}(b) and the discussion in the previous section.

For the intermediate cooling times, for which linear damping is more important than nonlinear dissipation ($\beta = 10^{-1}$ and $\beta = 1$; Fig.~\ref{fig:evolution}(c)--(d)), the resulting structure is dominated by a single wide gap around the orbit of the planet. In these cases, the gap extends from about $0.5 r_\mathrm{p}$ to about $1.5 r_\mathrm{p}$, so that its fractional width $\Delta r/r$ is $\approx 1$. Some hints of additional structure are present, but the multiple ring/gap structure is highly suppressed.

For the cases which develop multiple ring/gap structures, the features are more pronounced, i.e., the gaps are deeper, for short cooling timescales as compared to long cooling timescales. This is a consequence of the fact that for short cooling timescales (i.e., in locally isothermal or approximately locally isothermal disks), waves in the inner disk gain AMF from the background disk flow at larger radii and carry it to small radii, where it is returned to the disk through the nonlinear dissipation. This results in stronger features, as compared to disks with long cooling timescales, i.e., effectively adiabatic disks, for which waves do not gain AMF from the disk as they propagate, but can only dissipate their initial AMF. This difference in behaviors has been explored in \citet{Miranda-ALMA}.

\section{Discussion}
\label{sect:discussion}

\subsection{Implications for Protoplanetary Disks}

In protoplanetary disks, thermal relaxation is mediated by a combination of (i) radiative cooling via thermal dust emission from the surface of the disk and (ii) radiative diffusion along the plane of the disk. In general, the full energy evolution equation describing one or both of these processes may be linearized, resulting in a cooling law of the form given by equation~(\ref{eq:cooling_law}). The specific form of the resulting cooling timescale $t_\mathrm{c}$ or $\beta$ depends on the details of the cooling processes under consideration. Note that in general $\beta$ must vary with the distance from the star, unlike the idealized constant $\beta$ scenario considered in this work. However, our theoretical framework is not limited to the constant $\beta$ case, and can be applied to arbitrary $\beta(r)$ profiles.

In the case of radiative cooling from the surface of the disk, estimates of the cooling timescale $\sim e\Sigma/(\sigma T_\mathrm{eff}^4)$ (where $T_\mathrm{eff}$ is the disk effective temperature) for typical disk parameters lead to values of $\beta$ ranging from $\sim 100$ at several AU to $\sim 10^{-2}$ at $100$ AU (e.g., \citealt{Zhu2015}). We have shown (see Section \ref{sect:amf-numerical}) that in the linear regime, the locally isothermal approximation is applicable only for $\beta \lesssim 10^{-3}$. For more massive planets $\beta = 10^{-2}$ is only marginally compatible with the isothermal description, see \S \ref{sect:AMF_massive}. Based on this we conclude that even in the outer regions of protoplanetary disks, radiative cooling is not sufficiently rapid for the locally isothermal approximation to provide an accurate description of wave dynamics. 

However, in the optically thick inner disk (radii less than a few tens of AU), radiative diffusion may instead be the dominant source of cooling. In this case, the cooling timescale in equation~(\ref{eq:cooling_law}) should be identified with the diffusion timescale $(\eta k^2)^{-1}$, where $\eta$ is the radiative diffusion coefficient and $k$ is the radial wavenumber of the perturbation (e.g., \citealt{LinYoudin2015}). Note that in this case, the cooling timescale depends on the perturbation wavelength. In the context of the linear theory presented in this work, this translates to a dependence of the cooling timescale on the azimuthal mode number $m$. Rather than a universal cooling time, each Fourier harmonic of the internal energy is thermally relaxed on its own cooling timescale $t_{\mathrm{c},m}$.

It can be shown that the radiative diffusion cooling timescale is smaller than the (optically thick) radiative cooling timescale by a factor $(kH)^2 \sim m^2$ for a density wave with azimuthal number $m$. For planet-driven waves with a characteristic azimuthal number $\sim h_\mathrm{p}^{-1} \sim 10$, this suggests that the value of $\beta$ associated with radial diffusion may range from $\sim 1$ at several AU, down to some minimum value, perhaps $\sim 10^{-3}$, occuring at the optically thick/thin transition radius. Therefore, when cooling is mediated by thermal diffusion, the locally isothermal approximation may provide an accurate treatment of wave dynamics in the outermost regions of protoplanetary disks, but not at smaller radii, where a full consideration of the effects of cooling is indispensable. In a forthcoming work (Miranda \& Rafikov 2020, in prep.) we will explore the evolution of density waves using a detailed prescription for the cooling timescale, which accounts for both radiative cooling and diffusion, as appropriate for realistic protoplanetary disks.

\subsection{Implications for Multiple Spiral Arms}
\label{sect:discussion-spirals}

In Section~\ref{sect:pert_struct}, we pointed out several modifications of the multiple spiral arm structure of planet-driven density waves in disks with cooling. We now discuss the reasons behind these modifications in the context of AMF conservation.

As we noted, the basic fact that multiple spirals are formed in the inner disk is not modified by cooling. Multiple spiral arms are formed as a consequence of the mode interference governed by the dispersion relation for spiral density waves. This dispersion relation is not strongly modified by cooling (see equation~(\ref{eq:rek_cool})), with one key exception: the effective sound speed for density waves varies from $c_\mathrm{s,iso}$ for small values of $\beta$ to $c_\mathrm{s,adi}$ for large values of $\beta$. As a result, the pitch angle of the spirals gets effectively larger by a factor $\gamma^{1/2}$ (i.e., $18\%$ larger for $\gamma = 7/5$) for adiabatic disks as compared to locally isothermal disks, see Fig.~\ref{fig:spirals}(a),(e).

In regards to this fact, note that in \citet{Miranda-Spirals}, we pointed out that the spiral structure is nearly independent of the value of $\gamma$ for adiabatic disks. This statement applies when the comparison is made between disks which have the same adiabatic sound speed profile, i.e., for which $\gamma T(r)$ is the same. If instead we consider disks with different values of $\gamma$ but with the same temperature profile $T(r)$ (i.e., the same isothermal sound speed profile), then the spiral structure in the linear regime does in fact depend on the value of $\gamma$. Specifically, different values of $\gamma$ lead to different adiabatic sound speeds (proportional to $\gamma^{1/2}$). Correspondingly, the pitch angles of the spirals are larger, and the evolution of the single spiral into multiple spirals proceeds more slowly with distance from the planet, for larger values of $\gamma$ (and the same temperature profile). See Section~\ref{sect:cs} for a further discussion of this issue.

In \citet{Miranda-Spirals} we performed a detailed characterization of the spiral structure for adiabatic disks. However, with a minor modification it can also be applied to locally isothermal disks. Specifically, the quantity $\delta\Sigma_\mathrm{lin}$, which characterizes the overall scaling of the wave amplitude as a function of $r$ as dictated by AMF conservation (see equation~(16) of \citealt{Miranda-Spirals}), should be modified to account for the fact that $F_J$ is not constant in this case, but proportional to $c_\mathrm{s}^2$. Otherwise, the details of the emergence of the different spirals at different radii, their relative amplitudes and separations, and so on, are the same for both adiabatic and locally isothermal disks. This can be traced to the fact that the AMF for different wave modes $F_J^m$ all obey the same radial scaling in both cases: they are all constant in adiabatic disks, and all proportional to $c_\mathrm{s}^2$ in locally isothermal disks.

The situation is rather different in cooling disks with $\beta \approx 10^{-2}$ -- $1$. In such disks the radial scaling of $F_J^m$, and hence of the amplitudes of the harmonics of the surface density perturbation $\delta\Sigma_m$, varies with $m$, see Figs.~\ref{fig:amf_inner} and \ref{fig:amf_fourier}. As a result, the structure of the spirals is different than in either the short or long cooling timescale regimes. Therefore, the secondary spiral, tertiary spiral, and so on, form at different locations in the inner disk, and their relative amplitudes and widths evolve differently in these cases. For example, according to Fig.~\ref{fig:amf_inner}(a), in a disk with $\beta=0.1$ all the modes except for $m=2$ have small and positive $l^m_F/r$, implying rapid decay of these modes towards the inner disk. On the contrary, $m=2$ mode has large $l^m_F/r$ in a disk with this value of $\beta$. As a result, only the $m=2$ mode is present in the inner disk---note two broad, well azimuthally separated arms, clearly visible in Fig.~\ref{fig:spirals}(c) (see also the discussion in the end of Section \ref{sect:pert_struct}). We leave further quantitative analysis of the multiple spiral structure for disks with cooling to future work.

As a result of the modification of the spiral structure due to cooling in the linear regime, the subsequent nonlinear evolution of the spirals may also be modified. For example, if cooling reduces the amplitude of a spiral arm or broadens it, then it will develop into a shock at a larger distance from the planet (or potentially not at all). This results in a shift of the axisymmetric gap associated with the spiral developing into a shock. The modified shape of the spiral arm due to cooling changes the amount of AMF it carries and can ultimately deposit into the disk, therefore also affecting the depth of the resultant gap.

\subsection{Isothermal Versus Adiabatic Sound Speed}
\label{sect:cs}

It is important to distinguish between the different sound speeds relevant to wave dynamics in disks with different thermodynamics. These are the isothermal and adiabatic sound speeds:
\be
c_\mathrm{s,iso} = \left(\frac{k_\mathrm{B}T}{\mu}\right)^{1/2}, \ c_\mathrm{s,adi} = \left(\frac{\gamma k_\mathrm{B}T}{\mu}\right)^{1/2},
\ee
where $\mu$ is the mean molecular weight. Note that $c_\mathrm{s,adi} = \gamma^{1/2}c_\mathrm{s,iso}$. In Section~\ref{sect:theory}, we showed that $c_\mathrm{s,iso}$ is the appropriate sound speed in locally isothermal disks, while $c_\mathrm{s,adi}$ is the appropriate sound speed in adiabatic disks. We also showed that, with a finite cooling timescale, the appropriate sound speed is neither $c_\mathrm{s,adi}$ nor $c_\mathrm{s,iso}$, but can be represented by an effective sound speed intermediate between the two (see equation~(\ref{eq:cs-eff})). This effective sound speed is non-universal, with a non-trivial dependence on the cooling timescale, as well as on the location in the disk and the azimuthal number of the perturbation.

In numerical simulations, the disk temperature profile is typically prescribed indirectly through the sound speed, which itself is often parameterized by specifying the disk aspect ratio $h = H/r$, with the disk thickness $H = c_\mathrm{s}/\Omega$, where $c_\mathrm{s}$ is either the isothermal sound speed or the adiabatic sound speed. In locally isothermal simulations, $H$ must be defined in terms of $c_\mathrm{s,iso}$, since this is both the actual propagation speed of sound waves, and the only sound speed that is properly defined in this case. But in simulations with an ideal equation of state (with or without cooling), we can in principle choose to define $H$ in terms of either of the two sound speeds. This ambiguity is resolved by noting that, if $H$ is defined in terms of $c_\mathrm{s,iso}$, then disks with the same $h(r)$ profile have the same temperature profile, regardless of other thermodynamic considerations (adiabatic index or cooling timescale). If we regard the temperature as a fundamental physical property of the disk --- as opposed to the sound speed, which depends on thermodynamic assumptions---then this is the preferred way of defining $H$. Parameterizing the disk temperature in terms of $h = c_\mathrm{s,iso}/(r\Omega)$, as we have done in this paper, facilitates the most direct comparison of simulations of disks with different thermodynamics---locally isothermal, adiabatic, or cooling.

\subsection{Anomalous Mass Flux}
\label{sect:discussion-mdot}

\begin{figure}
\begin{center}
\includegraphics[width=0.49\textwidth,clip]{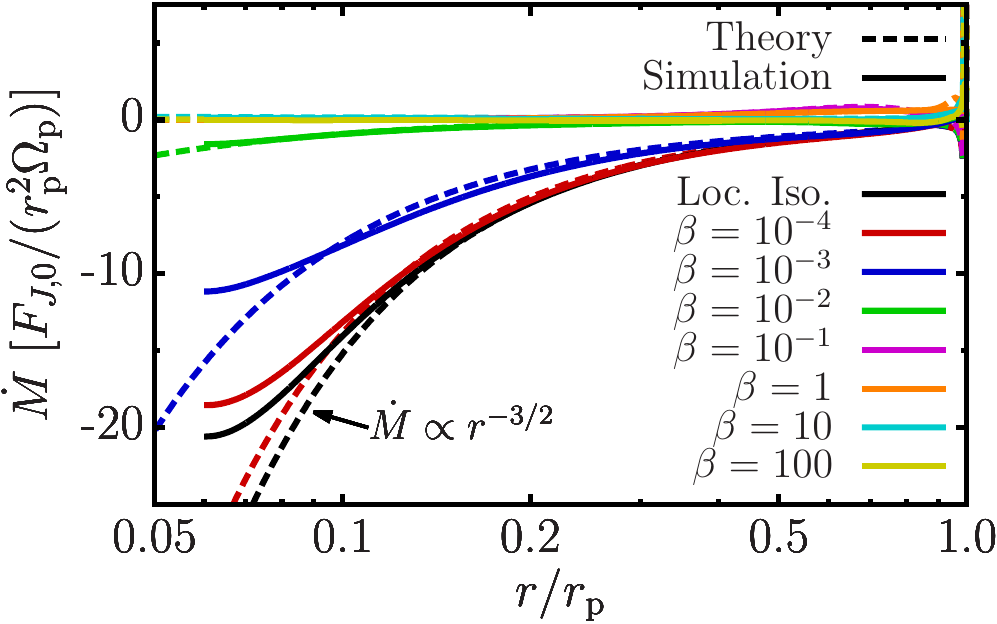}
\caption{Mass flux $\dot{M}$ due to planet-driven waves in the inner disk for disks with different dimensionless cooling timescales. The dashed lines show the theoretical mass flux (equation~(\ref{eq:mdot})) based on linear theory, and the solid lines show the results of numerical simulations (with $M_\mathrm{p} = 0.01 M_\mathrm{th}$). The asymptotic behavior for $r \ll r_\mathrm{p}$ in the locally isothermal limit (and for a $q = 1$ temperature profile) is indicated in the lower left.}
\label{fig:mdot}
\end{center}
\end{figure}

As a result of angular momentum conservation in the disk $+$ density wave system,
evolution of the wave AMF must result in disk evolution. In particular, the mass flux $\dot{M} = -\oint r\Sigma u_r \mathrm{d}\phi$ (here defined to be positive for inflow) is related to the evolution of the wave AMF according to \citep{R02b}
\be
\label{eq:mdot}
\dot{M} = \left(\frac{\mathrm{d}l}{\mathrm{d}r}\right)^{-1}\left(\frac{\mathrm{d}F_J}{\mathrm{d}r} - \frac{\mathrm{d}T}{\mathrm{d}r}\right),
\ee
where $l = r^2\Omega$ is the specific angular momentum. For simplicity we have neglected an additional term related to the time evolution of the angular momentum profile of the disk, although in some situations this term can be important (e.g., \citealt{AR18}). 

The physical interpretation of equation~(\ref{eq:mdot}) is as follows. The evolution of the wave AMF, described by $\mathrm{d}F_J/\mathrm{d}r$, occurs for two different reasons. The first is wave excitation---transfer of angular momentum to the wave by an external torque, described by $\mathrm{d}T/\mathrm{d}r$. Far from the planet this term can be neglected. The second is the transfer of angular momentum from the wave to the background disk (e.g. due to damping of the wave), which necessarily leads to disk evolution. Therefore, the disk evolution is determined by the difference between the total variation of the AMF and the external torque density, hence this difference appears in equation~(\ref{eq:mdot}).

Consider planet-driven waves which are subject to dissipation (either linear or nonlinear), so that $F_J$ decreases as waves propagate away from the planet. In this case, equation~(\ref{eq:mdot}) indicates that $\dot{M} < 0$ in the outer disk (as $\mathrm{d}F_J/\mathrm{d}r < 0$ there) and $\dot{M} >0$ in the inner disk (where $\mathrm{d}F_J/\mathrm{d}r > 0$). In other words, the effect of wave dissipation is effectively to {\it repel} mass from the orbit of the planet. In the absence of dissipation the wave has no effect on the state of the disk \citep{GN89}.

However, in locally isothermal disks, the wave AMF, described by equation~(\ref{eq:fj_law_iso}), increases as waves propagate towards the inner disk, and decreases in the outer disk (see Fig.~\ref{fig:amf_total}(a)), provided that the disk temperature decreases with $r$. As a result, in the locally isothermal case, using equation~(\ref{eq:mdot}) we find (setting $\mathrm{d}T/\mathrm{d}r = 0$ as applicable for free waves)
\be
\dot{M} = \frac{2 F_J}{r\Omega} \frac{\mathrm{d}\ln c_\mathrm{s}^2}{\mathrm{d}\ln r} = -\frac{2 q F_J}{r^2\Omega},
\ee
where the second equality applies to disks with $T \propto r^{-q}$. In this case, provided that $q > 0$, $\dot{M} < 0$ everywhere in the disk, with $|\dot{M}| \propto r^{-(q+1/2)}$ because $F_J\propto c_s^2\propto r^{-q}$. This means that in the inner disk there is a flow of mass {\it toward} the planet in the linear regime, before the waves shock, in contrast to the conventional expectation. In the outer disk, even before the wave shocks, there is a flow of mass {\it away} from the planetary orbit. This represents anomalous wave-driven disk evolution that is not associated with wave dissipation in the usual sense (cf. \citealt{GN89}).

The case of a disk with cooling is distinct from the adiabatic and locally isothermal cases: even in the absence of nonlinear dissipation, evolution of the wave flux due to cooling (see Fig.~\ref{fig:amf_total}(a)) will give rise to non-zero $\dot M$. However, depending on the value of $\beta$, $\dot M$ in the inner disk can be either positive (e.g., for intermediate values of $\beta \sim 0.1$) or negative (e.g., for small $\beta \lesssim 10^{-3}$, close to the locally isothermal limit).

The behavior of the anomalous mass flux described above is illustrated in Fig.~\ref{fig:mdot}. The mass flux in the inner disk from numerical simulations in the linear regime (for $M_\mathrm{p} = 0.01 M_\mathrm{th}$, when the wave does not shock) is shown along with the theoretical mass flux (equation~(\ref{eq:mdot})) predicted by linear theory (using $F_J$ computed in linear theory, see Fig.~\ref{fig:amf_total}(a)). The locally isothermal case and cases with cooling and different constant values of $\beta$ are shown. The numerical and theoretical results show good agreement, except at small radii ($r \lesssim 0.1 r_\mathrm{p}$).\footnote{Discrepancies at these small radii may be the result of nonlinear evolution of the waves, or boundary effects associated with the presence of the inner wave damping zone.} For $\beta \lesssim 10^{-2}$, as well as in the locally isothermal case, there is a substantial outward flow of gas, i.e., toward the planet ($\dot{M} < 0$). For $\beta = 10^{-1}$ -- $1$ one can see a small positive $\dot M$, as expected from the $F_J(r)$ behavior.

In Fig.~\ref{fig:mdot}, $\dot{M}$ is expressed in terms of $F_{J,0}/(r_\mathrm{p}^2\Omega_\mathrm{p})$ (see equation~(\ref{eq:fj0})). If we equate this scale to the mass flux for a steady state viscous disk, $3\pi \nu \Sigma$, with $\nu$ parameterized using the $\alpha$-prescription, then the characteristic scale of $\dot{M}$ due to planet-driven waves is equivalent to having 
\ba  
\alpha \sim \left(\frac{M_\mathrm{p}}{M_\mathrm{th}}\right)^2 h_\mathrm{p}.
\label{eq:alpha}
\ea  
For $M_\mathrm{p}\sim M_\mathrm{th}$ this characteristic $\alpha$ can substantially exceed the usual ``viscous'' $\alpha$.

However, in practice the aforementioned anomalous mass flux (away from the central star) can have a substantial impact on the disk evolution only when the wave amplitude is small enough for waves to travel far into the inner disk without shocking and nonlinear dissipation. This requires planets to be low mass with $M_\mathrm{p}$ below roughly a few percent of $M_\mathrm{th}$. In this case, the effective $\alpha$ would be $\lesssim 10^{-4}$. Nonlinear dissipation of the density waves driven by more massive ($M_\mathrm{p}\sim M_\mathrm{th}$) planets can give rise to substantially higher values of effective $\alpha$ \citep{GR01,R02}.

\subsection{Other Implications}
\label{sect:other}

It is commonly assumed that the locally isothermal approximation should be valid provided that $\beta = \Omega t_\mathrm{c} \lesssim 1$, i.e., the cooling timescale is short compared to the orbital timescale. However, we showed that this expectation is too naive and that, in fact, there are two key timescales to which $t_\mathrm{c}$ should be compared in order to assess the effects of cooling on the dynamics of density waves, neither of which are the orbital timescale.

The first relevant timescale is the time for a fluid element to cross through one of the crests/troughs of the perturbation with azimuthal number $m$, which is $|\tilde{\omega}|^{-1}$. This can be understood by noticing that $\beta$ enters equations (\ref{eq:c1_cool})--(\ref{eq:psi_cool}) only in the combination  $\tilde\beta=\tilde\omega t_\mathrm{c}=m\beta(\omega_\mathrm{p}-\Omega)/\Omega$. If the wave pattern frequency is slow, e.g., in the inner disk, then $|\tilde\beta|\approx m\beta$. When $|\tilde\beta| \sim 1$, there is a change in the behavior of the equations. Density waves experience strong linear damping when $|\tilde \beta| \sim 1$. Damping is minimal for $|\tilde\beta| \gg 1$, which represents the adiabatic limit, in which the wave AMF is constant. Damping is also weak when $|\tilde{\beta}| \ll 1$; however, this constraint, which is already more stringent than $\beta \lesssim 1$, still does not guarantee convergence to the locally isothermal limit, in which the AMF actually rises with the disk temperature.

This constraint is improved upon by considering the second relevant (dimensionless) timescale $\beta_\pm \sim h/m^2$ (equation~(\ref{eq:beta-pm})), for which the radial derivative of the AMF changes sign in the inner disk. Only when $\beta \lesssim \beta_\pm$, does the AMF grow as waves travel inwards. This represents a transition towards the locally isothermal regime, in which $F_J \propto r^{-q}$. Based on our numerical results, we require $\beta$ to be much smaller than $\beta_\pm$ (by at least an order of magnitude) for the AMF behavior to fully converge to the locally isothermal limit. For waves excited by planets, the dominant azimuthal number of perturbations is $m\sim h_\mathrm{p}^{-1} \sim 10$ \citep{GT80}, and so the locally isothermal limit corresponds to $\beta \lesssim h_\mathrm{p}^3 \sim 10^{-3}$ in realistic disks (see equation~(\ref{eq:beta-to-iso})). As waves become nonlinear, this requirement relaxes to $\beta \lesssim 10^{-2}$, see Section~\ref{sect:massive}. For other types of density waves (i.e., not driven by planets), the applicability of the locally isothermal approximation depends on their characteristic azimuthal wave number.

As a result of the stringent constraints on the cooling timescale necessary for the locally isothermal approximation to provide a good description of wave dynamics, its use is likely to be unjustified in many studies involving wave dynamics in disks. This is especially relevant for situations involving the long-range propagation of planet-excited density waves. These include modifications to disk-driven migration due to communication between planets via density waves \citep{Podlewska}, and due to wave reflection at disk edges \citep{Tsang2011,Miranda2018}. Under more realistic thermodynamics, the range over which the density waves can propagate may be limited, reducing the efficacy of these mechanisms. In studies of circumbinary accretion (e.g., \citealt{Munoz2019,Moody2019}), use of the locally isothermal approximation may lead to a misestimation of the size of the circumbinary cavity and of the circumstellar disks (or ``mini-disks'') around the individual stars. This could affect the balance of gravitational torques exerted on the binary by circumbinary and circumstellar disks, artificially influencing the resulting orbital evolution of the binary. 

A number of other problems critically dependent on globally propagating density waves have been studied using locally isothermal simulations. These include the origin of multiple narrow rings and gaps \citep{DongGaps2017,DongGaps2018,Bae2017,Nazari2019,Weber2019}, similar to the substructures seen in submillimeter observations of protoplanetary disks \citep{ALMA-HL,Andrews-TW,Isella2016,Loomis2017,Andrews-DSHARP,Long2018}, Papaloizou-Pringle instability \citep{Barker}, excitation of disk eccentricity in binaries \citep{Regaly2011}, and Lidov-Kozai oscillations in disks \citep{Martin2014}, among many others. It may be useful to reconsider some aspects of these problems in terms of AMF conservation as we do in our work. The impact of the locally isothermal EoS on the results of these studies, as well as any modifications of the results that may arise under the consideration of realistic thermodynamics should be identified.

The analysis presented in this work is 2D. We therefore expect its validity to depend on the extent to which the structure of  planet-excited waves is confined to the midplane of the disk. Irradiated disks may possess an increasing vertical temperature gradient. In the presence of such a temperature gradient, a portion of the angular momentum in planet-excited waves is carried by buoyancy waves, which are channeled toward the upper layers of the disk, possibly leading to enhanced nonlinear dissipation \citep{Lee2015}. It is unclear whether or not such an effect would be subdominant to the dissipation associated with cooling described in this work. However, our 2D analysis with cooling may set a lower bound to the amount of dissipation expected in 3D.

\section{Summary}
\label{eq:sum}

In this work we explored the properties of density waves in disks with varied thermodynamics, focusing on the effects of disk cooling on the wave propagation and damping. We used both linear theory and numerical simulations and used the behavior of the angular momentum flux (AMF) of the waves as a diagnostic of wave-driven disk evolution. Our main results can be summarized as follows.

\begin{itemize}

\item While the AMF of free waves $F_J$ is strictly conserved in adiabatic disks (in the absence of linear or nonlinear dissipation), it varies in locally isothermal disks as $F_J\propto c_\mathrm{s}^2$ (or the disk temperature $T$). 

\item In disks with a more general thermodynamics, in which the temperature is cooled/relaxed towards an equilibrium profile on a characteristic timescale $t_\mathrm{c}$, the adiabatic and locally isothermal limits are recovered when the cooling timescale is very long and very short, respectively. 

\item However, for the locally isothermal approximation to provide a good description of wave dynamics in the linear regime, the cooling timescale must be very short, with $\beta = \Omega t_\mathrm{c} \lesssim h_p^3\approx 10^{-3}$. 
This constraint relaxes to $\beta \lesssim 10^{-2}$ for more massive planets because of the nonlinear wave damping.

\item The adiabatic limit (i.e. conservation of $F_J$ in the linear regime) becomes applicable in disks with cooling for $\beta \gtrsim 10$. 

\item The transition between the two limiting regimes is non-monotonic and highly non-trivial. For intermediate cooling times, $\beta = 10^{-2}$ -- $1$, the wave AMF rapidly decays due to linear damping. 

\item Differences in the decay rates of different Fourier modes of the wave significantly modify the appearance of the planet-driven spiral structure in the inner disk for intermediate cooling times. 

\item Non-conservation of the wave AMF in locally isothermal and rapidly cooling disks gives rise to anomalous mass flux driving disk evolution even in the absence of viscosity or nonlinear dissipation.

\item In idealized disks with a radially constant dimensionless cooling time $\beta$, the structure of the gaps and rings carved in the disk by a moderately massive planet is strongly affected by the value of $\beta$. When $\beta$ is small ($\lesssim 10^{-1}$) or large ($\gtrsim 1$), multiple narrow rings and gaps (one ring/gap pair exterior to the orbit of the planet and several more interior to the orbit) are formed. For intermediate values of $\beta$, a single wide gap centered on the orbit of the planet is formed instead.

\end{itemize}

Our results should provide guidance for future efforts to better understand the appearance and evolution of density waves and to interpret observations of protoplanetary disks.

\acknowledgements

We are grateful to an anonymous referee for comments that helped improve the clarity of our presentation. Financial support for this work was provided by NASA via grant 15-XRP15-2-0139.

\appendix

\section{Perturbation Equations}
\label{sect:pert-eqns}

\subsection{Adiabatic}

For adiabatic disks, equations (\ref{eq:pert1})--(\ref{eq:pert3}) become
\begin{gather}
\label{eq:pert_adi1}
-\mathrm{i}\tilde{\omega}\frac{\Sigma}{c_\mathrm{s,adi}^2}\delta h + \frac{\Sigma}{L_S}\delta u_r+ \frac{1}{r}\frac{\partial}{\partial r}(r\Sigma\delta u_r) + \frac{\mathrm{i}m\Sigma}{r} \delta u_\phi = 0, \\
\label{eq:pert_adi2}
-\mathrm{i}\left(\frac{\tilde{\omega}^2-N_r^2}{\tilde{\omega}}\right)\delta u_r - 2\Omega\delta u_\phi = -\frac{\partial}{\partial r}(\delta h + \Phi_m) + \frac{\delta h}{L_S}, \\
\label{eq:pert_adi3}
-\mathrm{i}\tilde{\omega} \delta u_\phi + \frac{\kappa^2}{2\Omega} \delta u_r = -\frac{\mathrm{i}m}{r}\left(\delta h + \Phi_m\right).
\end{gather}

\subsection{Locally Isothermal}

For locally isothermal disks, equations (\ref{eq:pert1})--(\ref{eq:pert2}) become
\begin{gather}
\label{eq:pert_iso1}
-\mathrm{i}\tilde{\omega}\frac{\Sigma}{c_\mathrm{s,iso}^2}\delta h + \frac{1}{r}\frac{\partial}{\partial r}(r\Sigma\delta u_r) + \frac{\mathrm{i}m\Sigma}{r} \delta u_\phi = 0, \\
\label{eq:pert_iso2}
-\mathrm{i}\tilde{\omega}\delta u_r - 2\Omega\delta u_\phi = -\frac{\partial}{\partial r}(\delta h + \Phi_m) + \frac{\delta h}{L_T},
\end{gather}
and equation~(\ref{eq:pert3}) is the same as for adiabatic disks (\ref{eq:pert_adi3}).

\subsection{Disks with Cooling}

For disks with cooling, equations (\ref{eq:pert1})--(\ref{eq:pert2}) become
\begin{gather}
\begin{gathered}
\label{eq:pert_cool1}
\left(1+\frac{\mathrm{i}}{\gamma\tilde{\omega}t_\mathrm{c}}\right)^{-1} \left[\left(\frac{1}{t_\mathrm{c}}-\mathrm{i}\tilde{\omega}\right) \frac{\Sigma}{c_\mathrm{s,adi}^2} \delta h + \frac{\Sigma}{L_S}\delta u_r\right] \\ + \frac{1}{r}\frac{\partial}{\partial r} (r\Sigma\delta u_r) + \frac{\mathrm{i}m\Sigma}{r} \delta u_\phi = 0,
\end{gathered} \\
\begin{gathered}
\label{eq:pert_cool2}
-\mathrm{i}\left(\frac{\kappa^2-D_\mathrm{c}}{\tilde{\omega}}\right) \delta u_r - 2\Omega\delta u_\phi = -\frac{\partial}{\partial r} (\delta h + \Phi_m) \\ + \left(\frac{L_T^{-1}-\mathrm{i}\gamma\tilde{\omega}t_\mathrm{c}L_S^{-1}}{1-\mathrm{i}\gamma\tilde{\omega}t_\mathrm{c}}\right) \delta h,
\end{gathered}
\end{gather}
and again equation~(\ref{eq:pert3}) is the same as for adiabatic disks (\ref{eq:pert_adi3}).

\section{Velocity Perturbations}
\label{sect:velocity-pert}

\subsection{Adiabatic}

For adiabatic disks, the velocity perturbations are given in terms of $\delta h$ by
\begin{align}
\label{eq:ur_adi}
\delta u_r & = \frac{\mathrm{i}}{D_S}\left[\left(\tilde{\omega}\frac{\partial}{\partial r} - \frac{2m\Omega}{r}\right)(\delta h + \Phi_m) - \frac{\tilde{\omega}}{L_S}\delta h\right], \\
\begin{split}
\label{eq:uphi_adi}
\delta u_\phi & = \frac{1}{D_S}\left[\left(\frac{\kappa^2}{2\Omega}\frac{\partial}{\partial r} - \frac{m}{r}\left(\frac{\tilde{\omega}^2-N_r^2}{\tilde{\omega}}\right)\right)(\delta h + \Phi_m) \right. \\
& \left. - \frac{\kappa^2}{2\Omega L_S}\delta h\right].
\end{split}
\end{align}

\subsection{Locally Isothermal}

For locally isothermal disks, the velocity perturbations are given in terms of $\delta h$ by
\begin{align}
\label{eq:ur_iso}
\delta u_r & = \frac{\mathrm{i}}{D}\left[\left(\tilde{\omega}\frac{\partial}{\partial r} - \frac{2m\Omega}{r}\right)(\delta h + \Phi_m) - \frac{\tilde{\omega}}{L_T}\delta h\right], \\
\label{eq:uphi_iso}
\delta u_\phi & = \frac{1}{D}\left[\left(\frac{\kappa^2}{2\Omega}\frac{\partial}{\partial r} - \frac{m\tilde{\omega}}{r}\right)(\delta h + \Phi_m) - \frac{\kappa^2}{2\Omega L_T}\delta h\right].
\end{align}

\subsection{Disks with Cooling}

For disks with cooling, the velocity perturbations are given in terms of $\delta h$ by
\begin{align}
\label{eq:ur_cool}
\begin{split}
\delta u_r  & = \frac{\mathrm{i}}{D_\mathrm{c}} \Bigg[\left(\tilde{\omega} \frac{\partial}{\partial r} - \frac{2m\Omega}{r}\right)(\delta h + \Phi_m) \\
& - \left(\frac{L_T^{-1}-\mathrm{i}\gamma\tilde{\beta}L_S^{-1}}{1-\mathrm{i}\gamma\tilde{\beta}}\right)\tilde{\omega} \delta h \Bigg],
\end{split} \\
\label{eq:uphi_cool}
\begin{split}
\delta u_\phi & = \frac{1}{D\mathrm{c}}\Bigg[\left(\frac{\kappa^2}{2\Omega}\frac{\partial}{\partial r} - \frac{m}{r}\left(\frac{\kappa^2-D_\mathrm{c}}{\tilde{\omega}}\right)\right)(\delta h + \Phi_m) \\
& - \left(\frac{L_T^{-1}-\mathrm{i}\gamma\tilde{\beta}L_S^{-1}}{1-\mathrm{i}\gamma\tilde{\beta}}\right) \frac{\kappa^2}{2\Omega} \delta h\Bigg].
\end{split}
\end{align}

\section{WKB Analysis for Disks with Cooling}
\label{sect:wkb-cool}

Adopting the WKB ansatz, we write
\be
\delta h(r) = A(r) \exp\left[\mathrm{i}\int^r k(r^\prime)\mathrm{d}r^\prime\right],
\ee
where $k(r)$ is the radial wavenumber and $A(r)$ is a slowly varying amplitude. The master equation for free waves (equations (\ref{eq:master}), (\ref{eq:c1_cool})--(\ref{eq:psi_cool}) with $\Phi_m \rightarrow 0$) then reads
\be
\label{eq:master_wkb}
\frac{A^{\prime\prime}}{A} + \frac{2\mathrm{i}kA^\prime}{A} + \mathrm{i}k^\prime - k^2 + C_1 \left(\frac{A^\prime}{A} + \mathrm{i}k\right) + C_0 = 0.
\ee
Assuming $|kr| \gg 1$, to lowest order we have
\be
\label{eq:k2}
k^2 = - \left(\frac{1-\mathrm{i}\tilde{\beta}}{1-\mathrm{i}\gamma\tilde{\beta}}\right)\frac{\gamma D_\mathrm{c}}{c_\mathrm{s,adi}^2}.
\ee
In equation~(\ref{eq:k2}) we have assumed a thin disk, $c_\mathrm{s}/(r\Omega) \ll 1$, so that only the last term in equation~(\ref{eq:c0_cool}) for $C_0$ is retained. We see that in general $k$ is complex. We wish to obtain expressions for the real and imaginary parts of $k$ (rather than $k^2$), as we are interested in the attenuation coefficient $\mathrm{Im}(k)$. In doing so, we use the approximation $D_\mathrm{c} \approx D$, since $N_r^2$ is smaller than $D$ by $\mathcal{O}(h^2)$ (see equation~(\ref{eq:Dc})). We find
\be
\begin{aligned}
\label{eq:k}
k & = \gamma^{1/2} \left(\frac{1+\tilde{\beta}^2}{1+\gamma^2\tilde{\beta}^2}\right)^{1/4} \frac{|D|^{1/2}}{c_\mathrm{s,adi}} \\
& \times \exp\left\{\frac{\mathrm{i}}{2} \tan^{-1} \left[\frac{(\gamma-1)\tilde{\beta}}{1+\gamma\tilde{\beta}^2}\right]\right\}.
\end{aligned}
\ee
It can be shown that the argument of the inverse tangent function inside the exponential, $(\gamma-1)\tilde{\beta}/(1+\gamma\tilde{\beta}^2)$, is always small. For example, if $\gamma = 7/5$, its maximum possible (absolute) value is $\approx 0.17$. We therefore expand equation~(\ref{eq:k}) to leading order in this quantity and obtain equations (\ref{eq:imk_cool}) and (\ref{eq:rek_cool}) for the imaginary and real parts of $k$. 

\begin{figure}
\begin{center}
\includegraphics[width=0.49\textwidth,clip]{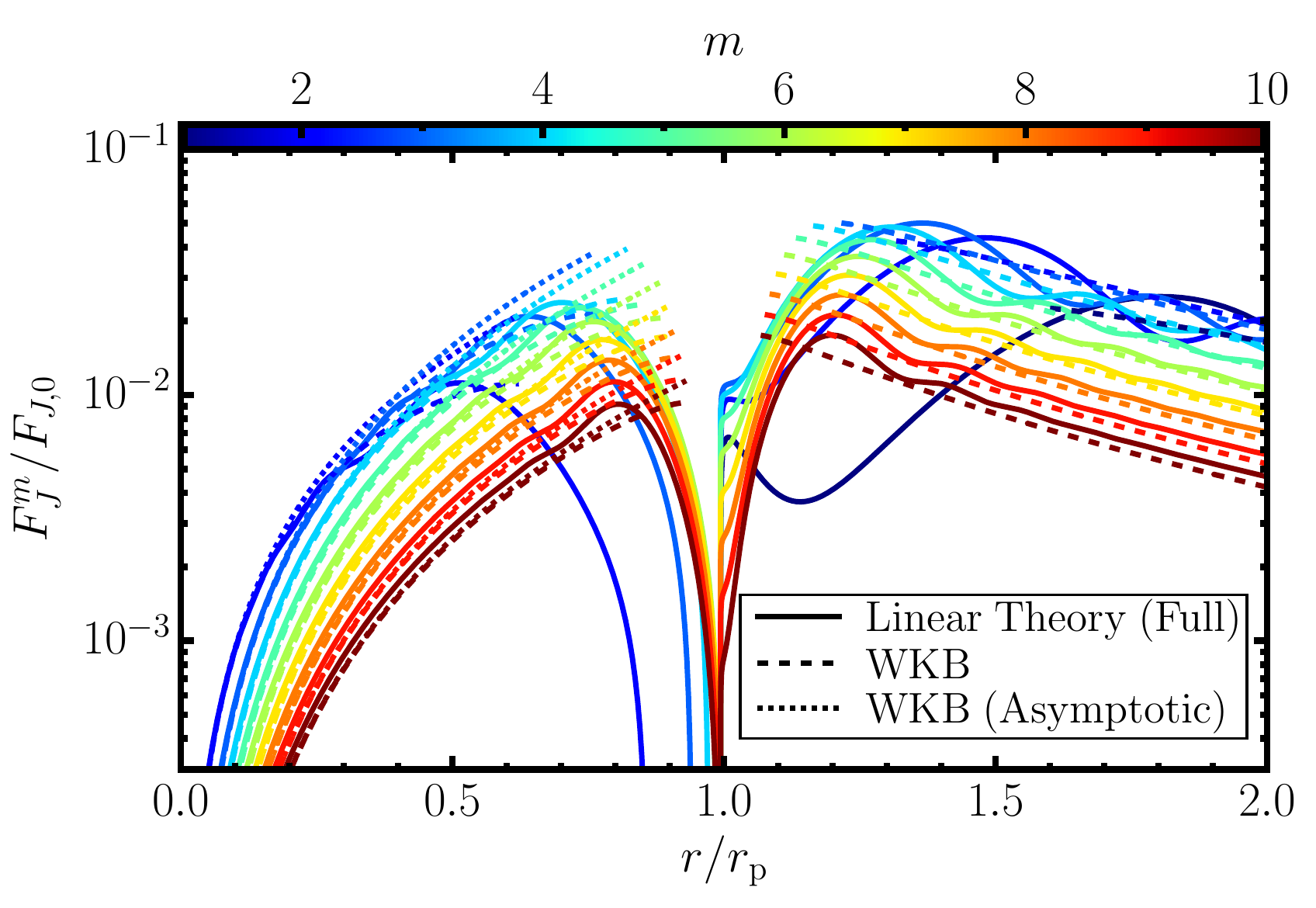}
\caption{Comparison of different approximations for $F_J^m$, the Fourier components of the wave AMF for the case with a dimensionless cooling timescale $\beta = 1$. The results of a fully self-consistent linear calculation are shown as solid lines, with the different colors corresponding to different azimuthal mode numbers $m$. The WKB approximation for $F_J^m$, (equation~(\ref{eq:fj_law_cool})), is represented by the dashed lines, and its asymptotic inner disk behavior (equation~(\ref{eq:fj_law_cool_inner})) by the dotted lines. The approximations for $F_J^m$ are only shown exterior to the outer Lindblad resonances and interior to the inner Lindblad resonances. They are also scaled by arbitrary amplitudes (as they are defined only up to an overall constant) to facilitate comparison with the full linear results.}
\label{fig:amf_fourier_wkb}
\end{center}
\end{figure}

To the next leading order ($|kr| \gg 1$) after equation (\ref{eq:k2}), equation~(\ref{eq:master_wkb}) gives
\be
\frac{2A^\prime}{A} + \frac{k^\prime}{k} + C_1 = 0,
\ee
and so the amplitude of $\delta h$ is
\be
\label{eq:amp}
A \propto \left(\frac{D}{r\Sigma k}\right)^{1/2} \exp\left[\frac{1}{2}\int^r \left(\frac{\Omega^2 +\mathrm{i}\gamma\Omega\tilde{\omega}\beta}{\Omega^2 + \gamma^2 \tilde{\omega}^2 \beta^2}\right) \frac{\mathrm{d}r^\prime}{L_T} \right],
\ee
where again we have taken $D_\mathrm{c} \approx D$.

\subsection{Angular Momentum Flux}

The AMF (equation~(\ref{eq:fj_cool})) is, using $|k\delta h| \gg |\delta h/r|$, and again using $D_S \approx D$ (or equivalently $D_\mathrm{c} \approx D$),
\be
\begin{aligned}
F_J^m & \approx \frac{\pi m r \Sigma}{D} \mathrm{Im}(\delta h \delta h^{*\prime}) \\
& \propto \frac{m r \Sigma}{D} \mathrm{Re}(k) |A|^2 \exp\left[-2 \int^r \mathrm{Im}(k) \mathrm{d}r^\prime\right].
\end{aligned}
\ee
Therefore, using (\ref{eq:imk_cool}) and (\ref{eq:amp}), we have
\be
\begin{aligned}
F_J^m(r) & = F_J^m(r_0) \\
& \times \exp\left\{\int_{r_0}^r \left[\frac{\Omega^2 L_T^{-1}}{\Omega^2 + \gamma^2 \tilde{\omega}^2 \beta^2} - 2 \mathrm{Im}(k)\right] \mathrm{d}r^\prime\right\},
\end{aligned}
\ee
where $r_0$ is an arbitrary reference radius. Note that we have taken a prefactor $\mathrm{Re}(k)/|k|$ in this expression to be $\approx 1$, since including $\mathrm{Im}(k)$ in this term gives only a small correction, and because $F_J$ depends on $\mathrm{Im}(k)$ much more strongly through the exponential.

\subsection{Validity of the WKB Approximation}
\label{sect:WKB-validate}

The validity of the WKB approximation of the Fourier components of the wave AMF (equation~(\ref{eq:fj_law_cool})) and its asymptotic inner disk approximation (equation~(\ref{eq:fj_law_cool_inner})) are examined in Fig.~\ref{fig:amf_fourier_wkb} for the case with $\beta = 1$. The WKB approximation (dashed lines in Fig.~\ref{fig:amf_fourier_wkb}) gives a good description of the radial variation of $F_J^m$ once the wave AMF has been fully accumulated, at distances from the planet $|r-r_\mathrm{p}|$ larger than about $2 |r_\mathrm{p}-r_\mathrm{LR}|$, where $r_\mathrm{LR}$ is the location of either the inner or outer Lindblad resonance (denoted by the endpoints of the dashed lines Fig.~\ref{fig:amf_fourier_wkb}). Therefore, it is typically valid beyond $2$--$3$ scale heights from the planet, except for modes with small azimuthal numbers ($m \lesssim 2$). The inner disk WKB approximation (short dashed lines in Fig.~\ref{fig:amf_fourier_wkb}) has approximately the same region of validity.

\bibliographystyle{apj}
\bibliography{references}

\begin{thebibliography}{}
\expandafter\ifx\csname natexlab\endcsname\relax\def\natexlab#1{#1}\fi

\bibitem[{{ALMA Partnership} {et~al.}(2015){ALMA Partnership}, {Brogan},
  {P{\'e}rez}, {Hunter}, {Dent}, {Hales}, {Hills}, {Corder}, {Fomalont},
  {Vlahakis}, {Asaki}, {Barkats}, {Hirota}, {Hodge}, {Impellizzeri}, {Kneissl},
  {Liuzzo}, {Lucas}, {Marcelino}, {Matsushita}, {Nakanishi}, {Phillips},
  {Richards}, {Toledo}, {Aladro}, {Broguiere}, {Cortes}, {Cortes}, {Espada},
  {Galarza}, {Garcia-Appadoo}, {Guzman-Ramirez}, {Humphreys}, {Jung}, {Kameno},
  {Laing}, {Leon}, {Marconi}, {Mignano}, {Nikolic}, {Nyman}, {Radiszcz},
  {Remijan}, {Rod{\'o}n}, {Sawada}, {Takahashi}, {Tilanus}, {Vila Vilaro},
  {Watson}, {Wiklind}, {Akiyama}, {Chapillon}, {de Gregorio-Monsalvo}, {Di
  Francesco}, {Gueth}, {Kawamura}, {Lee}, {Nguyen Luong}, {Mangum}, {Pietu},
  {Sanhueza}, {Saigo}, {Takakuwa}, {Ubach}, {van Kempen}, {Wootten},
  {Castro-Carrizo}, {Francke}, {Gallardo}, {Garcia}, {Gonzalez}, {Hill},
  {Kaminski}, {Kurono}, {Liu}, {Lopez}, {Morales}, {Plarre}, {Schieven},
  {Testi}, {Videla}, {Villard}, {Andreani}, {Hibbard}, \&
  {Tatematsu}}]{ALMA-HL}
{ALMA Partnership}, {Brogan}, C.~L., {P{\'e}rez}, L.~M., {et~al.} 2015, \apjl,
  808, L3

\bibitem[{{Andrews} {et~al.}(2016){Andrews}, {Wilner}, {Zhu}, {Birnstiel},
  {Carpenter}, {P{\'e}rez}, {Bai}, {{\"O}berg}, {Hughes}, {Isella}, \&
  {Ricci}}]{Andrews-TW}
{Andrews}, S.~M., {Wilner}, D.~J., {Zhu}, Z., {et~al.} 2016, \apjl, 820, L40

\bibitem[{{Andrews} {et~al.}(2018){Andrews}, {Huang}, {P{\'e}rez}, {Isella},
  {Dullemond}, {Kurtovic}, {Guzm{\'a}n}, {Carpenter}, {Wilner}, {Zhang}, {Zhu},
  {Birnstiel}, {Bai}, {Benisty}, {Hughes}, {{\"O}berg}, \&
  {Ricci}}]{Andrews-DSHARP}
{Andrews}, S.~M., {Huang}, J., {P{\'e}rez}, L.~M., {et~al.} 2018, \apjl, 869,
  L41

\bibitem[{{Arzamasskiy} \& {Rafikov}(2018)}]{AR18}
{Arzamasskiy}, L., \& {Rafikov}, R.~R. 2018, \apj, 854, 84

\bibitem[{{Bae} \& {Zhu}(2018)}]{BZ18a}
{Bae}, J., \& {Zhu}, Z. 2018, \apj, 859, 118

\bibitem[{{Bae} {et~al.}(2017){Bae}, {Zhu}, \& {Hartmann}}]{Bae2017}
{Bae}, J., {Zhu}, Z., \& {Hartmann}, L. 2017, \apj, 850, 201

\bibitem[{{Barker} \& {Ogilvie}(2016)}]{Barker}
{Barker}, A.~J., \& {Ogilvie}, G.~I. 2016, \mnras, 458, 3739

\bibitem[{{Baruteau} \& {Masset}(2008)}]{Baruteau2008}
{Baruteau}, C., \& {Masset}, F. 2008, \apj, 672, 1054

\bibitem[{{Ben{\'{\i}}tez-Llambay} \& {Masset}(2016)}]{FARGO3d}
{Ben{\'{\i}}tez-Llambay}, P., \& {Masset}, F.~S. 2016, \apjs, 223, 11

\bibitem[{{de Val-Borro} {et~al.}(2006){de Val-Borro}, {Edgar}, {Artymowicz},
  {Ciecielag}, {Cresswell}, {D'Angelo}, {Delgado-Donate}, {Dirksen}, {Fromang},
  {Gawryszczak}, {Klahr}, {Kley}, {Lyra}, {Masset}, {Mellema}, {Nelson},
  {Paardekooper}, {Peplinski}, {Pierens}, {Plewa}, {Rice}, {Sch{\"a}fer}, \&
  {Speith}}]{deValBorro2006}
{de Val-Borro}, M., {Edgar}, R.~G., {Artymowicz}, P., {et~al.} 2006, \mnras,
  370, 529

\bibitem[{{Dong} {et~al.}(2017){Dong}, {Li}, {Chiang}, \& {Li}}]{DongGaps2017}
{Dong}, R., {Li}, S., {Chiang}, E., \& {Li}, H. 2017, \apj, 843, 127

\bibitem[{{Dong} {et~al.}(2018){Dong}, {Li}, {Chiang}, \& {Li}}]{DongGaps2018}
---. 2018, \apj, 866, 110

\bibitem[{{Dong} {et~al.}(2011{\natexlab{a}}){Dong}, {Rafikov}, \&
  {Stone}}]{Dong2011b}
{Dong}, R., {Rafikov}, R.~R., \& {Stone}, J.~M. 2011{\natexlab{a}}, \apj, 741,
  57

\bibitem[{{Dong} {et~al.}(2011{\natexlab{b}}){Dong}, {Rafikov}, {Stone}, \&
  {Petrovich}}]{Dong2011a}
{Dong}, R., {Rafikov}, R.~R., {Stone}, J.~M., \& {Petrovich}, C.
  2011{\natexlab{b}}, \apj, 741, 56

\bibitem[{{Goldreich} \& {Nicholson}(1989)}]{GN89}
{Goldreich}, P., \& {Nicholson}, P.~D. 1989, \apj, 342, 1075

\bibitem[{{Goldreich} \& {Tremaine}(1979)}]{GT79}
{Goldreich}, P., \& {Tremaine}, S. 1979, \apj, 233, 857

\bibitem[{{Goldreich} \& {Tremaine}(1980)}]{GT80}
---. 1980, \apj, 241, 425

\bibitem[{{Goodman} \& {Rafikov}(2001)}]{GR01}
{Goodman}, J., \& {Rafikov}, R.~R. 2001, \apj, 552, 793

\bibitem[{{Isella} {et~al.}(2016){Isella}, {Guidi}, {Testi}, {Liu}, {Li}, {Li},
  {Weaver}, {Boehler}, {Carperter}, {De Gregorio-Monsalvo}, {Manara}, {Natta},
  {P{\'e}rez}, {Ricci}, {Sargent}, {Tazzari}, \& {Turner}}]{Isella2016}
{Isella}, A., {Guidi}, G., {Testi}, L., {et~al.} 2016, Physical Review Letters,
  117, 251101

\bibitem[{{Korycansky} \& {Pollack}(1993)}]{KP93}
{Korycansky}, D.~G., \& {Pollack}, J.~B. 1993, \icarus, 102, 150

\bibitem[{{Lee}(2016)}]{Lee2016}
{Lee}, W.-K. 2016, \apj, 832, 166

\bibitem[{{Lee} \& {Gu}(2015)}]{Lee2015}
{Lee}, W.-K., \& {Gu}, P.-G. 2015, \apj, 814, 72

\bibitem[{{Lin}(2015)}]{Lin2015}
{Lin}, M.-K. 2015, \mnras, 448, 3806

\bibitem[{{Lin} \& {Papaloizou}(2011)}]{Lin2011}
{Lin}, M.-K., \& {Papaloizou}, J.~C.~B. 2011, \mnras, 415, 1445

\bibitem[{{Lin} \& {Youdin}(2015)}]{LinYoudin2015}
{Lin}, M.-K., \& {Youdin}, A.~N. 2015, \apj, 811, 17

\bibitem[{{Long} {et~al.}(2018){Long}, {Pinilla}, {Herczeg}, {Harsono},
  {Dipierro}, {Pascucci}, {Hendler}, {Tazzari}, {Ragusa}, {Salyk}, {Edwards},
  {Lodato}, {van de Plas}, {Johnstone}, {Liu}, {Boehler}, {Cabrit}, {Manara},
  {Menard}, {Mulders}, {Nisini}, {Fischer}, {Rigliaco}, {Banzatti}, {Avenhaus},
  \& {Gully-Santiago}}]{Long2018}
{Long}, F., {Pinilla}, P., {Herczeg}, G.~J., {et~al.} 2018, The Astrophysical
  Journal, 869, 17

\bibitem[{{Loomis} {et~al.}(2017){Loomis}, {{\"O}berg}, {Andrews}, \&
  {MacGregor}}]{Loomis2017}
{Loomis}, R.~A., {{\"O}berg}, K.~I., {Andrews}, S.~M., \& {MacGregor}, M.~A.
  2017, \apj, 840, 23

\bibitem[{{Lunine} \& {Stevenson}(1982)}]{Lunine1982}
{Lunine}, J.~I., \& {Stevenson}, D.~J. 1982, \icarus, 52, 14

\bibitem[{{Martin} {et~al.}(2014){Martin}, {Nixon}, {Lubow}, {Armitage},
  {Price}, {Do{\u g}an}, \& {King}}]{Martin2014}
{Martin}, R.~G., {Nixon}, C., {Lubow}, S.~H., {et~al.} 2014, \apjl, 792, L33

\bibitem[{{Miranda} \& {Lai}(2018)}]{Miranda2018}
{Miranda}, R., \& {Lai}, D. 2018, \mnras, 473, 5267

\bibitem[{{Miranda} \& {Rafikov}(2019{\natexlab{a}})}]{Miranda-Spirals}
{Miranda}, R., \& {Rafikov}, R.~R. 2019{\natexlab{a}}, \apj, 875, 37

\bibitem[{{Miranda} \& {Rafikov}(2019{\natexlab{b}})}]{Miranda-ALMA}
---. 2019{\natexlab{b}}, The Astrophysical Journal, 878, L9

\bibitem[{{Moody} {et~al.}(2019){Moody}, {Shi}, \& {Stone}}]{Moody2019}
{Moody}, M. S.~L., {Shi}, J.-M., \& {Stone}, J.~M. 2019, \apj, 875, 66

\bibitem[{{Mu{\~n}oz} {et~al.}(2019){Mu{\~n}oz}, {Miranda}, \&
  {Lai}}]{Munoz2019}
{Mu{\~n}oz}, D.~J., {Miranda}, R., \& {Lai}, D. 2019, \apj, 871, 84

\bibitem[{{Nazari} {et~al.}(2019){Nazari}, {Booth}, {Clarke}, {Rosotti},
  {Tazzari}, {Juhasz}, \& {Meru}}]{Nazari2019}
{Nazari}, P., {Booth}, R.~A., {Clarke}, C.~J., {et~al.} 2019, \mnras, 485, 5914

\bibitem[{{Ogilvie} \& {Lubow}(2002)}]{OL02}
{Ogilvie}, G.~I., \& {Lubow}, S.~H. 2002, \mnras, 330, 950

\bibitem[{{Paardekooper} {et~al.}(2010){Paardekooper}, {Baruteau}, {Crida}, \&
  {Kley}}]{Paardekooper2010}
{Paardekooper}, S.~J., {Baruteau}, C., {Crida}, A., \& {Kley}, W. 2010, \mnras,
  401, 1950

\bibitem[{{Paardekooper} \& {Papaloizou}(2008)}]{Paardekooper2008}
{Paardekooper}, S.~J., \& {Papaloizou}, J.~C.~B. 2008, \aap, 485, 877

\bibitem[{{Papaloizou} {et~al.}(2007){Papaloizou}, {Nelson}, {Kley}, {Masset},
  \& {Artymowicz}}]{Papaloizou2007}
{Papaloizou}, J.~C.~B., {Nelson}, R.~P., {Kley}, W., {Masset}, F.~S., \&
  {Artymowicz}, P. 2007, in Protostars and Planets V, ed. B.~{Reipurth},
  D.~{Jewitt}, \& K.~{Keil}, 655

\bibitem[{{Podlewska-Gaca} {et~al.}(2012){Podlewska-Gaca}, {Papaloizou}, \&
  {Szuszkiewicz}}]{Podlewska}
{Podlewska-Gaca}, E., {Papaloizou}, J.~C.~B., \& {Szuszkiewicz}, E. 2012,
  \mnras, 421, 1736

\bibitem[{{Rafikov}(2002{\natexlab{a}})}]{R02}
{Rafikov}, R.~R. 2002{\natexlab{a}}, \apj, 569, 997

\bibitem[{{Rafikov}(2002{\natexlab{b}})}]{R02b}
---. 2002{\natexlab{b}}, \apj, 572, 566

\bibitem[{{Rafikov}(2016)}]{R16}
---. 2016, \apj, 831, 122

\bibitem[{{Rafikov} \& {Petrovich}(2012)}]{RP12}
{Rafikov}, R.~R., \& {Petrovich}, C. 2012, \apj, 747, 24

\bibitem[{{Reg{\'a}ly} {et~al.}(2011){Reg{\'a}ly}, {S{\'a}ndor}, {Dullemond},
  \& {Kiss}}]{Regaly2011}
{Reg{\'a}ly}, Z., {S{\'a}ndor}, Z., {Dullemond}, C.~P., \& {Kiss}, L.~L. 2011,
  \aap, 528, A93

\bibitem[{{Takeuchi} {et~al.}(1996){Takeuchi}, {Miyama}, \&
  {Lin}}]{Takeuchi1996}
{Takeuchi}, T., {Miyama}, S.~M., \& {Lin}, D.~N.~C. 1996, \apj, 460, 832

\bibitem[{{Tsang}(2011)}]{Tsang2011}
{Tsang}, D. 2011, \apj, 741, 109

\bibitem[{{Tsang}(2014)}]{Tsang2014}
---. 2014, \apj, 782, 112

\bibitem[{{Ward}(1997)}]{Ward1997}
{Ward}, W.~R. 1997, \icarus, 126, 261

\bibitem[{{Weber} {et~al.}(2019){Weber}, {P{\'e}rez}, {Ben{\'\i}tez-Llambay},
  {Gressel}, {Casassus}, \& {Krapp}}]{Weber2019}
{Weber}, P., {P{\'e}rez}, S., {Ben{\'\i}tez-Llambay}, P., {et~al.} 2019, \apj,
  884, 178

\bibitem[{{Zhang} \& {Lai}(2006)}]{Zhang2006}
{Zhang}, H., \& {Lai}, D. 2006, \mnras, 368, 917

\bibitem[{{Zhu} {et~al.}(2015){Zhu}, {Dong}, {Stone}, \& {Rafikov}}]{Zhu2015}
{Zhu}, Z., {Dong}, R., {Stone}, J.~M., \& {Rafikov}, R.~R. 2015, \apj, 813, 88

\end{thebibliography}

\end{document}